\documentclass[11pt,british]{article}
\usepackage[utf8]{inputenc}
\usepackage[T1]{fontenc}
\usepackage{mathtools}
\usepackage{amsmath}
\usepackage{amsthm}
\usepackage{amssymb}
\usepackage{microtype}
\usepackage{Gmath}
\usepackage{babel}
\usepackage{stmaryrd}
\usepackage{slashed}
\usepackage{cancel}
\usepackage{centernot}
\usepackage{bbm}
\usepackage{stmaryrd}

\usepackage{import}
\usepackage{tikz}
\usepackage{graphicx}
\usepackage{comment}
\usepackage{tikz-cd}

\makeatletter
\usepackage{tikz}
\usepackage{tikz-cd}
\usetikzlibrary{backgrounds}
\usepackage[framemethod=tikz]{mdframed}
\AtBeginEnvironment{mdframed}{%
\tikzset{every picture/.style={}}%
}
\mdfsetup{roundcorner=.5ex}

\theoremstyle{definition}
\newtheorem*{defn*}{Definition}

\usepackage{jheppub}                   
\usepackage[linecolor=blue,backgroundcolor=blue!25,bordercolor=blue,textsize=scriptsize]{todonotes}

\makeatletter
\gdef\@fpheader{\ }                    
\makeatother

\usepackage{setspace}
\usepackage{amsfonts} 		
\SetTracking{encoding={*}, shape=sc}{0} 	
\usepackage{color} 		
\definecolor{darkblue}{rgb}{0.0,0.0,0.3} 	
\allowdisplaybreaks		
\date{\today} 		
\numberwithin{equation}{section}	

\makeatletter
\g@addto@macro\bfseries{\boldmath}
\makeatother

\let\originalleft\left
\let\originalright\right
\renewcommand{\left}{\mathopen{}\mathclose\bgroup\originalleft}
\renewcommand{\right}{\aftergroup\egroup\originalright}

\newcommand{\DW}[1]{\textbf{DW:\ #1}} 

\title{M-theory Moduli from Exceptional
Complex Structures}
\author[a]{George Robert Smith}
\emailAdd{g.smith19@imperial.ac.uk}
\author[a]{and Daniel Waldram}
\emailAdd{d.waldram@imperial.ac.uk}

\affiliation[a]{Department of Physics,
	Imperial College London, \\
	Prince Consort Road, London, SW7 2AZ, UK} 

\subheader{\hfill\textrm{Imperial/TP/22/DW/1}}

\abstract{
We continue the analysis of the geometry of generic Minkowski $\mathcal{N}=1$, $D=4$ flux compactifications in M-theory using exceptional generalised geometry, including the calculation of the infinitesimal moduli spaces. The backgrounds can be classified into two classes: type-0 and type-3. For type-0, we review how the moduli arise from standard de Rham cohomology classes. We also argue that, under reasonable assumptions, there are no appropriate sources to support compact flux backgrounds for this class and so the only solutions are in fact $\G_2$ geometries. For type-3 backgrounds, given a suitable $\del'\delb'$-lemma, we show that the moduli can be calculated from a cohomology based on an involutive sub-bundle of the complexified tangent space. Using a simple spectral sequence we prove quite generally that the presence of flux can only reduce the number of moduli compared with the fluxless case. We then use the formalism to calculate the moduli of heterotic M-theory and show they match those of the dual Hull--Strominger system as expected.}

\begin{document}
\maketitle

\section{Introduction}\label{sec:introduction}

This paper is the offspring of~\cite{1910.04795}, where generic flux compactifications of ten- and eleven-dimensional supergravity to four-dimensional Minkowski space preserving $\mathcal{N}=1$ supersymmetry were analysed using exceptional generalised geometry. The geometries have a simple interpretation as integrable $\SU{7}\subset \ExR{7(7)}$ generalised G-structures. In eleven-dimensional supergravity, these come in two classes, type-0 and type-3, and the earlier paper calculated several properties of the type-0 backgrounds, including their infinitesimal moduli spaces.

The main focus of this paper is to extend the analysis to the type-3 class and to discuss in some more detail the structure of the infinitesimal moduli spaces that appear. This is in principle directly relevant to model-building~\cite{Acharya:1998pm,Acharya:2000gb,Witten:2001uq,Atiyah:2001qf,Acharya:2001gy,Beasley:2002db,Berglund:2002hw,Acharya:2002kv,Atiyah:2001qf,DallAgata:2005zlf,Grigorian:2008tc,Acharya:2008hi,Braun:2018vhk}, where fluxes are typically included to try and lift some of the moduli fields. The usual analysis considers parametrically small fluxes perturbing a purely geometrical supersymmetric geometry, for example a $\G_2$-holonomy background. One of the powers of using exceptional generalised geometry techniques is that the moduli can be calculated even when the fluxes are large. 

There is a long history of analysing the M-theory $\mathcal{N}=1$ backgrounds using conventional G-structure techniques~\cite{Kaste:2003zd,Behrndt:2003zg,DallAgata:2003txk,Lukas:2004ip}. The advantage of using exceptional generalised geometry is that the Killing spinor equations are neatly rephrased into vanishing intrinsic torsion conditions for generalised G-structures~\cite{0804.1362,1112.3989,1606.09304,1910.04795} which readily incorporate the inclusion of fluxes.  This in turn provides a framework that is amenable to moduli problems on backgrounds with and without flux. In particular, moduli still correspond to elements of cohomology groups and these cohomology groups can be readily found as components of a deformation of the exceptional generalised G-structure. For the type-0 class the relevant groups are de Rham cohomologies. For the type-3 class, we will find that particular seven-dimensional analogues of Dolbeault cohomology groups appear.  In both cases we are simply deforming the generalised G-structure. 

There is a famous no-go theorem for M-theory supersymmeric compactifications on a space $M$ to Minkowski space~\cite{hep-th/0007018}. If $M$ is compact then there are no solutions with non-zero flux unless there are extra negative tension sources. Typically these come from M5-branes and orbifold planes in the geometry, the heterotic M-theory background~\cite{hep-th/9710208,hep-th/9806051} in Ho\v{r}ava--Witten theory~\cite{hep-th/9510209,hep-th/9603142} being the prime example. Our backgrounds will be assumed to have sources of this type, which are localised and carry some net overall charge, even if smeared. The main result of this paper then comes in two parts:
\begin{enumerate}
    \item supersymmetric backgrounds with non-trivial flux are type-3,
    \item type-3 moduli can be counted by cohomology groups using a three page spectral sequence, such that the number of moduli weakly decreases with the inclusion of flux.
\end{enumerate}
Thus even without the assumption that the flux is a small perturbation, they can only reduce the number of moduli when compared to the pure metric geometry. 

The layout of the paper is as follows. In section \ref{sec:Exceptional complex structures} we first give a brief review of the Killing spinors and spinor bilinears following~\cite{Lukas:2004ip}, introduce exceptional complex geometry and the notion of an $\SU7$ structure and then relate the two approaches.  In section~\ref{sec:Integrability} we review the results of~\cite{1910.04795}, compute the integrability conditions for an $\SU{7}$ structure for type-0 and type-3 backgrounds and relate these results to the differential equations given in~\cite{Lukas:2004ip}.  It is in this section that we show that only type-3 structures can include flux.  In section \ref{sec:Deformations} we compute the infinitesimal moduli of the $\SU{7}$ structures, and show that type-3 structure moduli come from a simple spectral sequence, where the flux can obstruct certain deformations.  In section \ref{sec:Heterotic M-Theory} we compute the type-3 spectral sequence for the heterotic M-theory geometry $M_7=X\times S^1/\bbZ_2$, finding that the moduli match those calculated for the dual Hull--Strominger~\cite{Hull:1986kz,Strominger:1986uh} heterotic backgrounds in \cite{Anderson:2014xha,delaOssa:2014cia,delaOssa:2015maa,Ashmore:2019rkx}.

\section{Exceptional complex structures}\label{sec:Exceptional complex structures}

We are interested in eleven-dimensional supergravity compactifications to four-dimensional Minkowski space that preserve $\mathcal{N}=1$ supersymmetry. In order to preserve Lorentz invariance the eleven dimensional space is a warped product of Minkowski space $\bbR^{3,1}$ with an internal space $M$.  The eleven dimensional metric is
\begin{equation}\label{def:warped-metric}
    \dd s^2 = \ee^{2\Delta}\dd s^2(\bbR^{3,1}) + \dd s^2(M)
\end{equation}
and the eleven-dimensional flux takes the form $\mathcal{F} = F + \star \tilde{F}$
where $F$ and $\tilde{F}$ are a four-form and seven-form on $M$ respectively and $\star$ is the Hodge star operator. The corresponding Bianchi identities read 
\begin{equation}
    \dd F=\dd \tilde{F}+\tfrac12 F\wedge F=0
\end{equation}
so that one can introduce the corresponding local potentials
\begin{equation}
    F = \dd A , \qquad 
    \tilde{F} = \dd \tilde{A} - \tfrac12 A \wedge F . 
\end{equation}
In this section, we start by recalling the analysis of such backgrounds using conventional G-structure techniques following Lukas and Saffin \cite{Lukas:2004ip}. We then show how these conventional structures are in one-to-one correspondence with (almost) exceptional complex structures (ECS) in generalised geometry. In addition, using the explicit map between the conventional and generalised structures, we show that ECSs come in two ``types'', analogous to the ``type'' of generalised complex structures introduced by Gualtieri \cite{Gualtieri:2003dx}. 

\subsection{Killing Spinors and their Bilinears}
Generic eleven dimensional supergravity compactifications which preserve $\mathcal{N}=1$ supersymmetry have been analysed using Killing spinors and G-structures for decades \cite{hep-th/0303127,hep-th/0311146,hep-th/0311119,hep-th/0401019,hep-th/0205050,hep-th/0302158,hep-th/0212008}. First we summarise the analysis of Lukas and Saffin \cite{Lukas:2004ip}, specialised to the case of vanishing four-dimensional cosmological constant.  

The most general form for the eleven-dimensional Killing spinor\footnote{Our spinor conventions follow those of \cite{hep-th/0403235,1212.1586}. In particular $\gamma=\ii \gamma_{(4)}=\ii\gamma^0\gamma^1\gamma^2\gamma^3$.} is
\begin{equation}\label{def:Killing-spinor-ansatz}
    \eta = \theta\otimes \epsilon_{(1)} + \ii \gamma \theta\otimes\epsilon_{(2)}.
\end{equation}
where  the four-dimensional Majorana spinor $\theta$ is constant, while the real $\Spin{7}$ Majorana spinors $\epsilon_{(i)}$ are free to vary over the internal space. The latter define a complex $\Spin{7}$ spinor $\zeta_-$ and its conjugate 
\begin{equation}
    \zeta_{+} = \epsilon_1 + \ii\epsilon_2, \quad 
    \zeta_{-} = \zeta_+^c = \epsilon_1 - \ii\epsilon_2 .
\end{equation}
The decomposition uniquely defines $\zeta_+$ and $\theta$ up to a constant chiral rotation of $\theta$, under which $\zeta_+\mapsto\ee^{\ii\alpha}\zeta_+$, and a constant real rescaling $\theta\mapsto\mu^{-1}\theta$ and $\zeta_+\mapsto\mu\zeta_+$. One can use the former freedom to set $\epsilon_{(1)}^T \epsilon_{(2)}=0$ and in addition choose $|\epsilon_{(2)}|\leq|\epsilon_{(1)}|$, so that we can parameterise the ratio of the norms as 
\begin{equation}\label{def:L&S_chi_def}
  \frac{|\epsilon_2|}{|\epsilon_1|} = \frac{\cos\chi}{1+\sin\chi},
\end{equation}
At a point on $M$, the pair of spinors are generically stabilised by an $\SU{3}\subset\Spin7$ subgroup and so define a local $\SU3$ structure. However, at points where $\cos\chi=0$ we only have a single real spinor and so the structure group becomes $\Gx2$. A local $\SU3$ structure in seven dimensions may be defined via a set of three invariant forms $\{\sigma, \omega, \Omega\}$. The one-form $\sigma$ encodes a product structure such that the metric on $M$ decomposes 
\begin{equation}
\label{eq:metric-SU3}
    \dd s^2(M) = \dd s^2_{\SU3} + \sigma^2 , 
\end{equation}
while the real two-form $\omega$ and the complex three-form $\Omega = \Omega_+ + \ii \Omega_-$, define the six-dimensional $\SU3$ structure metric $\dd s^2_{\SU3}$.  One can us $\{\zeta_+, \zeta_-\}$ to construct a set of spinor bilinears, as found in Appendix B of \cite{Lukas:2004ip}\footnote{Note that the condition  $\epsilon_{(1)}^T \epsilon_{(2)}=0$ means that we can write $\tilde{\Xi}=\zeta_+^\dagger\zeta_-$ as $\tilde{\Xi}=\zeta_-^\dagger\zeta_+$}:
\begin{subequations}
\label{def:Lukas-and-saffin-appendix-equations}
\begin{align}
    \Xi &= \zeta_+^\dagger\zeta_+ = \ee^\Delta , \label{def:Lukas-and-saffin-appendix-equation-a}\\
    \tilde{\Xi} &= \zeta_-^\dagger\zeta_+ = \ee^\Delta\sin\chi ,  \label{def:Lukas-and-saffin-appendix-equation-b}\\
    \Xi_{(1)} &= - \zeta_+^\dagger \gamma_{(1)} \zeta_+ = \ee^\Delta \cos\chi\, \sigma ,  \label{def:Lukas-and-saffin-appendix-equation-c}\\
    \Xi_{(2)} &= i\zeta_+^\dagger \gamma_{(2)}\zeta_+ = \ee^\Delta\cos\chi\, \omega ,  \label{def:Lukas-and-saffin-appendix-equation-d}\\
    \Xi_{(3)} &= i\zeta_+^\dagger \gamma_{(3)}\zeta_+ = \ee^\Delta\Big( \sin\chi\, \Omega_- - \sigma\wedge\omega\Big), \label{def:Lukas-and-saffin-appendix-equation-e}\\
    \tilde{\Xi}_{(3)} &= -\zeta_-^\dagger \gamma_{(3)}\zeta_+ =\ee^\Delta\Big( \cos\chi\,  \Omega_+ + \ii\big(\Omega_- - \sin\chi\, \sigma\wedge\omega\big)\Big),  \label{def:Lukas-and-saffin-appendix-equation-f}
\end{align}
\end{subequations}
The bilinears are stabilised by the same structure which stabilises the pair of Killing spinors. In particular, when $\cos \chi=0$ we see that $\Xi_{(1)}$ and $\Xi_{(2)}$ vanish and $\Xi_{(3)}=\pm\ii\tilde{\Xi}_{(3)}=\varphi$ where $\varphi$ is the $\Gx2$-invariant three-form. Note that in all cases $\tilXi{}$ and $\tilXi{3}$ completely determine the $\SU3$ or $\Gx2$ structure, and hence the metric and $\Xi$, $\Xi_{(1)}$, $\Xi_{(2)}$ and $\Xi_{(3)}$ can be written in terms of (in general complicated) non-linear functions of $\tilXi{}$ and $\tilXi{3}$.  

At this point the factor $\ee^\Delta$ is an arbitrary function, in principle unrelated to the warp factor in the metric. However, translated into conditions on the bilinears, the internal Killing spinor equations imply 
\begin{align*}
    \dd(\ee^{-\Delta}\Xi) = 0 , 
\end{align*}
and so, using the constant scaling symmetry we may solve to find $\Xi=\ee^\Delta$, justifying the identification made above. The remaining spinor bilinears satisfy
\begin{subequations}
\label{eq:Lukas-and-saffin-differential-equations}
\begin{align}
    \dd(\ee^{2\Delta}\tilXi{}) &= 0 ,  \label{eq:Lukas-and-saffin-differential-equation-a}\\
    \dd(\ee^{\Delta}\tensXi{1}) &= 0 ,  \label{eq:Lukas-and-saffin-differential-equation-b}\\
    \dd(\ee^{3\Delta}\tensXi{2}) &= -\ee^{4\Delta}\star F ,  \label{eq:Lukas-and-saffin-differential-equation-c}\\
    \dd(\ee^{2\Delta}\tilXi{(3)}) &= - \ee^{2\Delta}\tilde{\Xi} F ,   \label{eq:Lukas-and-saffin-differential-equation-d}
\end{align}
\end{subequations}
where now $\star$ is the seven-dimensional Hodge star operator.  The external Killing spinor equations, which have been used to eliminate $\tilde{F}$ from \eqref{eq:Lukas-and-saffin-differential-equations}, can be written as
\begin{subequations}\label{eq:Lukas-and-saffin-differential-equations2}
\begin{align}
    \tilde{\Xi} \tilde{F} &= - \tfrac12 F \wedge \tilXi{(3)} , \label{eq:Lukas-and-saffin-differential-equation2a}\\
    \tilXi{(3)} \lrcorner F &= \ee^{4\Delta} \dd \tilXi{} . 
    \label{eq:Lukas-and-saffin-differential-equation2b}
\end{align}
\end{subequations}
As discussed in~\cite{Lukas:2004ip}, together these equations imply $\tilde{F} = 0$. Note that these conditions do not prevent pointwise vanishing of spinors, that is points where $\cos\chi=0$, and so the structure group may vary between $\Gx{2}$ and $SU(3)\subset \Gx{2}$ over the internal space $M$ and there is no sense of a proper (global) G-structure. We will see that this is only possible in cases where $\chi\neq 0$, where the flux is necessarily trivial.  With non-trivial flux we will see that $\chi=0$ globally, and we have a global $\SU{3}$ structure.

\subsection{$\cN=1$ supersymmetry and exceptional complex structures}

It was shown in \cite{1112.3989,1910.04795} that generic four-dimensional $\mathcal{N}=1$ Minkowski compactifciations are in one-to-one correspondence with integrable $\SU{7}$ structures in Exceptional Generalised Geometry. To see how this construction works let us briefly review the construction of the $\ExR{7(7)}$ geometry. 

The generalised vector bundle $E$ corresponding to the real $\repsub{56}{1}$-dimensional $\ExR{7(7)}$ representation can be locally decomposed in conventional tensor bundles as
\begin{equation}\label{def:generalised_vector}
\begin{aligned}
    E &\simeq T\oplus \ext^2 T^* \oplus \ext^5 T^*\oplus (T^*\otimes \ext^7 T^*) , \\
    \repsub{56}{1} &= \rep7 \oplus \rep{21} \oplus \rep{21'} \oplus \rep{7'}
\end{aligned}
\end{equation}
where we give the decomposition under $\SL{7,\bbR}\subset\ExR{7(7)}$. The subscript denotes the $\bbR^+$ weight, where $\det T^*$ transforms with a weight of $+2$. Other $\ExR{7(7)}$ representation bundles we will use are the adjoint bundle transforming in $\repsub{133}{0}\oplus\repsub{1}{0}$, with a $\GL{7,\bbR}$ decomposition \cite{1112.3989,1212.1586}:
\begin{equation}\label{def:generalised_adjoint}
\begin{aligned}
    \ad \tilde{F} &\simeq \bbR\oplus (T\otimes T^*)\oplus \ext^3 T^*\oplus \ext^6 T^* \oplus \ext ^3 T \oplus \ext^6 T , \\
    \repsub{133}{0} \oplus \repsub{1}{0} &= \rep{1} \oplus (\rep{48} \oplus \rep{1}) \oplus \rep{35} \oplus \rep{7} \oplus \rep{35'} \oplus \rep{7'} , 
\end{aligned}
\end{equation}
the sub-bundle of the symmetric product $S^2E\supset N\simeq (\det T^*)\otimes\ad\tilde{F}$, transforming as $\repsub{133}{2}$, so that the leading terms have the form
\begin{equation}\label{def:generlised_symmetric_adjoint}
    N \simeq T^*\oplus \ext^4 T^*\oplus (T^*\otimes \ext^6T^*)\oplus \dots ,
\end{equation}
and finally a bundle in the $\repsub{912}{3}$ representation~\cite{1910.04795,1112.3989} 
\begin{equation}
\begin{aligned}
    (\det T^*)^2\otimes K &\simeq \bbR \oplus \ext^3 T^*\oplus (T^*\otimes \ext^5 T^*) \oplus (S^2T^*\otimes \ext^7T^*)\oplus  \dots \\
    \repsub{912}{3} &= \rep{1} \oplus \rep{35} \oplus (\rep{140} \oplus \rep{7}) \oplus \rep{28} \oplus \dots
\end{aligned}
\end{equation}
where $K$ above is the bundle containing the generalised torsion and transforming in the $\repsub{912}{-1}$ $\ExR{7(7)}$ representation. 

The internal bosonic supergravity fields $\{g,A,\tilde{A}\}$ on $M$ together with the warp factor $\Delta$ encode a generalised metric $\mathcal{G}\in\Gamma(S^2E^*)$ which defines a generalised $\SU{8}/\mathbb{Z}_2$-structure \cite{0804.1362} (where $\SU{8}/\bbZ_2$ is the maximal compact subgroup of $\ExR{7(7)}$).  Furthermore the spinor fields and supersymmetry parameters form representations of the corresponding double-cover ``spinor'' group $\SU8$. In particular, the complex spinor $\zeta_+$ transforms in the fundamental $\rep{8}$ representation.  We immediately see that the supersymmetric background fixed by specifying the generalise metric $\mathcal{G}$ and the spinor $\zeta_+$ defines a generalised $G$-structure with 
\begin{equation}
    G = \Stab(\mathcal{G},\zeta_+) = \SU7 . 
\end{equation}
where $\Stab(\mathcal{G},\zeta_+)$ is the subgroup of $\ExR{7(7)}$ that leaves the pair $(\mathcal{G},\zeta_+)$ invariant. It is useful to decompose $\ExR{7(7)}$ representations for  the bundle $E$, $\ad\tilde{F}$ and $K$ under the $\SU7$ structure group. We find 
\begin{alignat}{2}
    \rep{56} &= \rep{28} \oplus \text{c.c.} 
        &&= \repsub{7}{3} \oplus \repsub{21}{-1} \oplus \text{c.c.} , 
    \label{eq:56-decompositions} \\
    \rep{133} &= \rep{63} \oplus \left(\rep{35} \oplus \text{c.c} \right) 
        &&= \repsub{1}{0} \oplus \repsub{48}{0} \oplus \left( \repsub{7}{-4} \oplus \repsub{35}{2} \oplus \text{c.c} \right) , 
    \label{eq:133-decompositions} \\
    \rep{912} &= \rep{36} \oplus \rep{420} \oplus \text{c.c} 
        &&= \repsub{1}{7} \oplus \repsub{7}{3} \oplus \repsub{21}{-1} \oplus \repsub{28}{-1} \oplus \repsub{35}{-5} \oplus \repsub{140}{3}\oplus \repsub{244}{-1} \oplus \text{c.c.} , 
    \label{eq:912-decompositions}
\end{alignat}
where we first give the decomposition under $\SU8$, and the subscript denotes the charge under the $\Uni1\subset\SU8$ that commutes with $\SU7$.  

There is a real singlet in the $\rep{133}$ representation and a complex singlet in the $\rep{912}$ representations, corresponding to generalised stabilised by an $\SU{7}$ structure \cite{1910.04795}.  Denote these tensors as
\begin{equation}
    J \in \Gamma(\ad\tilde{F}) , \qquad
    \psi \in \Gamma((\det T^*)^2 \otimes K_\bbC) .
\end{equation}
The stabiliser groups of each tensor satisfies the following embeddings:
\begin{equation*}
    \begin{tikzcd}[ampersand replacement=\&]
        \substack{\SU{7}\\\text{stab } \psi}\arrow[r, hook]\arrow[drr, hook] \& \substack{\Uni{7}\times \bbR^+\\\text{stab } J}\arrow[r, hook] \& \frac{\SU{8}}{\mathbb{Z}_2}\times\bbR^+\arrow[r, hook] \& \ExR{7(7)}\\
         \&  \& \substack{\SU{8}/\mathbb{Z}_2\\\text{stab }\mathcal{G}} \arrow[ur, hook]\&
    \end{tikzcd}
\end{equation*}
We see that $J$ is invariant under a $\Uni{7}\times\bbR^+$ subgroup of $\ExR{7(7)}$. The $\bbR^+$ is not a subgroup of $\SU{8}/\bbZ_2$ so specifying $J$ determines $\mathcal{G}$ up to a scale, given by the warp factor. On the other had  $\psi$ is stabilised by a unique $SU(7)$ group which is a subgroup of $\SU{8}/\bbZ_2$, and specifying $\psi$ gives a unique $J$ and $\mathcal{G}$. 

We call the generalised tensor $J$ an (almost) exceptional complex structure. It inherits the adjoint action on the fibres of the complexified vector bundle $E_\bbC$ and generates the $\Uni1\subset\SU8/\bbZ_2$ that commutes with the $\SU7$ structure. Under this action the fibres at each point decompose into four eigenbundles  of the operator $J$ with eigenvalues $\{\pm 3i,\pm i\}$ corresponding to the second decomposition in~\eqref{eq:56-decompositions} 
\begin{align*}
    E_\bbC = L_3 \oplus L_1\oplus L_{-1} \oplus L_{-3}.
\end{align*}
In fact, specifying a suitable $L_3$ sub-bundle is equivalent to specifying $J$ at each point. Following \cite{1910.04795}, the \textit{type} of a $\UniR{7}$ or $\SU{7}\subset \UniR{7}$ structure is defined at a point $p\in M$ to be
\begin{equation}\label{def:type}
    \text{type } L_3 = \text{codim } \pi \big(L_3\big) = 7 - \dim \pi \big(L_3\big).
\end{equation}
where the anchor map $\pi:E \to T$ is the projection onto the $T_\bbC M$ tangent space component of $E_\bbC$. Note that, a priori, the type of $L_3$ can change as one moves over the manifold $M$.  At a point $p\in M$ we can view the choice of $J$ or $\psi$ as a element in the relevant coset spaces
\begin{equation}
\label{eq:cosets}
    \left. J\right|_p \in \frac{\Ex{7(7)}}{\Uni7} , \qquad
    \left. \psi \right|_p \in \frac{\ExR{7(7)}}{\SU7} ,
\end{equation}
and any two exceptional complex structures can be related to each other by an $\ExR{7(7)}$ transformation. These spaces are 84- and 86-dimensional respectively. 

\subsection{Spinor bilinears and exceptional complex structure type}

Having abstractly defined exceptional complex structures, we now show how they can be explicitly written in terms of the bilinears defined in~\eqref{def:Lukas-and-saffin-appendix-equations}. As discussed in~\cite{1910.04795}, the $J$ and $\psi$ singlets live in the $\rep{63}$ and $\rep{36}$ representations of $\SU8$ respectively and can be written explicitly in terms of the spinor $\zeta_+$ as 
\begin{equation}
    J = \zeta_+\zeta^\dag_+ - \tfrac{1}{8}(\zeta^\dag_+\zeta_+) \id , 
    \qquad
    \psi = \zeta_+\zeta_+^T = \zeta_+\zeta^\dag_- . 
\end{equation}
Comparing with~\eqref{def:Lukas-and-saffin-appendix-equations} we see that $J$ and should depend linearly on the bilinears $\Xi_{(1)}$, $\Xi_{(2)}$ and $\Xi_{(3)}$, while $\psi$ depends on $\tilde{\Xi}$ and $\tilde{\Xi}_{(3)}$.  Explicitly
\begin{align}
\label{def:ECC_from_spinor_bilinears}
    J & = \ee^{A+\tilde{A}}\cdot\Big(- \Xi_{(2)R} +\Xi_{(3)} - \star\Xi_{(1)} 
        - \Xi^\sharp_{(3)} - \big(\star\Xi_{(1)}\big)^\sharp\Big) , \\
    \psi & = \ee^{A+\tilde{A}}\cdot\ee^{2\Delta} \Big(\tilde{\Xi} + \tilde{\Xi}_{(3)}         
        + jg\wedge\star\tilde{\Xi}_{(3)} + \tilde{\Xi}\, (g\otimes \vol_g) + \dots \Big) , 
\end{align}
where we are using the notation $\Xi_{(2)R}$ for $(\Xi_{(2)})^m{}_n\in\Gamma(T\otimes T^*)$, $\Xi^\sharp_{(3)}$ for $(\Xi^\sharp_{(3)})^{mnp}\in\Gamma(\ext^3T)$ etc, and $ jg\wedge\star\tilde{\Xi}_{(3)}$ for $5g_{m[(n_1}(\star\tilde{\Xi}_{(3)})_{n_2\dots n_5}\in\Gamma(T^*\otimes\ext^5 T^*)$, and the potential $\ee^{A+\tilde{A}}$ factors denote the exponentiated $\Ex{7(7)}$ adjoint action. 

From the form of $\psi$ we immediately see that $\sin\chi=0$ corresponds to the special case where $\tilXi{}$ vanishes, and the leading term in $\psi$ is a three-form rather than a scalar. In fact we can rewrite $\psi$ in the following form in the two cases
\begin{equation}\label{eq:psi_from_spinor_bilinears}
\psi = 
    \begin{cases}
        \ee^{A+\tilde{A}}\,\ee^{\tilXi{3}/\tilXi{}}\cdot \ee^{2\Delta}\tilde{\Xi} , & \quad \tilde{\Xi}\neq0 \\
        \ee^{A+\tilde{A}}\,\ee^{-\ii\sigma\wedge\omega - \frac{1}{8}\Omega\wedge\Bar{\Omega}}\cdot \ee^{2\Delta}\Omega , &
        \quad \tilde{\Xi} = 0 . 
    \end{cases}
\end{equation}
We can also solve the eigenvalue equation to find $L_3$ in terms of the spinor bilinears, from which we can read off the type. One finds
\begin{align}\label{eq:L_3_from_spinor_bilinears}
L_3 = 
    \begin{cases}
        \ee^{A+\tilde{A}}\,\ee^{\tilXi{3}/\tilXi{}}\cdot T_\bbC, & \tilXi{} \neq 0,  \quad \text{type-0} \\
        \ee^{A+\tilde{A}}\,\ee^{-\ii\sigma\wedge\omega - \frac{1}{8}\Omega\wedge\Bar{\Omega}}\cdot \big(\ker\Omega \oplus \image \Omega), & \tilXi{} = 0, \quad \text{type-3}.
    \end{cases}
\end{align}
where we are viewing $\Omega$ as a map $\Omega:T_\bbC\to\ext^2T^*_\bbC$ given by $v\mapsto\imath_v\Omega$, so that $\dim(\ker\Omega)=4$ and $\dim(\im\Omega)=3$. Note that there are precisely two types, type-0 and type-3, and the spinor bilinear $\tilXi{}$ distinguishes between them.  Notice also that despite $L_3$ defining type above and there being a clear distinction between types this is harder to see from the form of $J$.  We can view type-3 structures as the $\chi\to 0$ limit of tpye 0 structures, and we use this view to calculate the above expressions for type-3 structures in Appendix \ref{app:type-3 as a limit of type-0}. Finally we see that type-0 is the generic structure with type-3 of measure zero in the space of structures, corresponding to setting $\tilde{\Xi}=0$.

In \eqref{eq:L_3_from_spinor_bilinears} the degrees of freedom in specifying $L_3$ are packaged into the spinor bilinears. However there is enough freedom in the generic $\SU3$ structure to write a type-0 almost exceptional complex structure as
\begin{align}\label{def:type_0_original}
    L_3 = \ee^{\alpha+\beta}T_\bbC
\end{align}
where $\alpha\in \Lambda^3T^*_{\bbC}$ and $\beta\in \Lambda^6T^*_\bbC$ are almost generic forms.  In particular, they parameterise 84 real degrees of freedom corresponding to the dimension of the coset space of $J$ structures in \eqref{eq:cosets}. The conditions given in~\cite{1910.04795} on $L_3$ for it to correspond to an eigenbundle of $J$ reduce to the requirement that $\im \alpha$ is a positive three-form in the sense of Hitchin \cite{arXiv:math/0107101}. This is an open condition on $\im \alpha$ so has no effect on the number of degrees of freedom. In the special case that $\beta$ vanishes and $\alpha$ is purely imaginary the bundle $L_3\to M$ defines a $\Gx{2}$ structure on $M$ with the stabilised non-degenerate 3-form $\im \alpha$, so $\Gx{2}$ compactifications are contained within the space of type-0 structures. Given the general relations 
\begin{align}
\label{eq:ab-blinear-0}
   \alpha= A + \tilXi{(3)}/\tilXi, \qqq \beta = \tilde{A} - \tfrac{1}{2}A\wedge\big(\tilXi{(3)}/\tilXi{ }\big)
\end{align}
the combination
\begin{equation}
\label{eq:6form-combo}
    \im \beta + \tfrac{1}{2}\re \alpha \wedge \im \alpha = \frac{\cos \chi}{2\sin^2\chi}\Omega_+\wedge\Omega_-
\end{equation}
is nonzero and independent of changing the gauge fields. It produces a well defined $6$-form on $M$ breaking the $\Gx{2}$ structure defined by $\im \alpha$ to an $\SU{3}$ structure. We thus see, without directly referring to the spinor bilinears that a type-0 structure generically defines a $\SU{3}$ structure with $\Gx{2}$ as a special case when $\cos\chi=0$.

For the type-3 structure, the underlying conventional structure defined by the spinor bilinears is always $\SU3$ and we can always decompose the complexified tangent space as 
\begin{equation}
    T_\bbC \simeq T^{(1,0)} \oplus T^{(0,1)} \oplus \bbC \sigma^\sharp
\end{equation}
writing the corresponding vector components as $v^m=(v^a,v^{\bar{a}},v^7)$. We can then identify
\begin{equation}
\label{eq:T01-def}
    T^{[0,1]} := \ker\Omega \simeq  T^{(0,1)} \oplus \bbC \sigma^\sharp , \qquad
    T^{[1,0]} := T_\bbC/T^{[0,1]}  \simeq  T^{(1,0)} ,
\end{equation}
and define the corresponding $[p,q]$ forms 
\begin{equation}
\label{eq:Lambdapq-def}
    \Lambda^{[p,q]} = \ext^pT^{[1,0]*} \otimes \ext^qT^{[0,1]*} \simeq \Lambda^{(p,q)} \oplus \Lambda^{(p,q-1)}\wedge \sigma , 
\end{equation}
where $\Lambda^{(p,q)}=\ext^pT^{(1,0)*} \otimes \ext^qT^{(0,1)*}$. If only the $\Omega$ structure is specified one can identify $T^{[0,1]}$ and $\Lambda^{[p,0]}$ as sub-bundles of $T_\bbC$ and $\ext^pT^*_\bbC$, respectively, such that, in particular 
\begin{align}
    \ker\Omega = T^{[0,1]}\qqq \text{ and } \qqq \im\Omega = \Lambda^{[2,0]} ,
\end{align}
while all the other bundles are formally defined as equivalence classes. However, using the full $\SU3$ structure we get a natural way to also identify $T^{[1,0]}$ and $\Lambda^{[p,q]}$ as subbundles, as in~\eqref{eq:T01-def} and~\eqref{eq:Lambdapq-def}. A generic type-3 structure can then be written as 
\begin{equation}
\label{eq:gen-type3-L}
    L_3 = \ee^{a+b} \cdot \big( T^{[0,1]} \oplus \Lambda^{[2,0]} \big) , 
    \qquad a \in \Gamma(\Lambda^{[1,2]}\oplus\Lambda^{[0,3]}) , 
    \quad b \in \Gamma(\Lambda^6T^*_\bbC) ,
\end{equation}
since the action of $a\in\Gamma(\Lambda^{[2,1]}\oplus\Lambda^{[3,0]})$ does not change $T^{[0,1]} \oplus \Lambda^{[2,0]}$. Counting degrees of freedom, we note that $T^{[0,1]}$ defines a $\GL{3, \bbC}\times \bbR \ltimes \bbC^3\subset\GL{7,\bbR}$ structure, where the stabilizer group is generated by  $\{r^{\bar{a}}{}_{\bar{b}}, r^7{}_{\bar{b}}, r^7{}_7\}\in\gl{7,\bbR}$. The space of such structures is $49-25=24$ dimensional, while $a$ and $b$ parametrise 58 degrees of freedom giving a total of 82, reflecting the fact that the subspace $\sin\chi=0$ is real codimension-two in the space of ECSs. The combination~\eqref{eq:6form-combo} has to be non-vansishing and defines the sub-bundle $\bbR\sigma^\sharp\subset T$ as the annihilator of the six-form. This breaks the structure to $\GL{3,\bbC}\times\bbR$ and gives an explicit identification $T^{[0,1]}\simeq T^{(0,1)}\oplus\bbC\sigma^\sharp$. Using the corresponding identification for $\Lambda^{[p,q]}$ given in~\eqref{eq:Lambdapq-def}, one notes that the only self-conjugate sub-bundle in $\Lambda^{[1,2]}\oplus\Lambda^{[0,3]}$ is $\Lambda^{(1,1)}\wedge\sigma$. This means that, up to terms that annihilate $ T^{[0,1]} \oplus \Lambda^{[2,0]}$,
we can always write $a$ and $b$ in the form 
\begin{equation}
   \alpha= A - \ii\sigma\wedge\omega , \qquad 
   \beta = \Tilde{A}-\tfrac{1}{8}\Omega\wedge\bar{\Omega}
    +\tfrac{1}{2}\ii A\wedge\sigma\wedge\omega  
\end{equation}
hence deriving the bilinear expression~\eqref{eq:L_3_from_spinor_bilinears}.

\section{Integrability}\label{sec:Integrability}

As first shown in~\cite{1112.3989} and discussed in detail in~\cite{1910.04795},  four-dimensional $\mathcal{N}= 1$ supersymmetric  warped Minkowski compactifications are in one-to-one correspondence with torsion-free generalised $\SU{7}$-structures on $M$. Torsion-free means that the structure admits a compatible connection with vanishing torsion, which, in general, puts differential constraints on the structure tensor $\psi$. The equivalence means that these differential constraints on $\psi$ exactly reproduce the conditions on the spinor bilinears given in~\eqref{eq:Lukas-and-saffin-differential-equations}. In this section we start by briefly summarising the results of~\cite{1910.04795} and then show how they apply to type-0 and type-3 structures. 

The obstruction to finding a torsion-free compatible connection is measured by the intrinsic torsion. This is the part of the torsion that is independent of the choice of compatible connection. For the $\SU{7}$ structure it is a section of a sub-bundle $W^{\text{int}}_{\SU{7}}$ of the $\repsub{912}{1}$ bundle $K$ which contains only a few $\SU{7}$ irreps, while for $\Uni7\times\bbR^+$ structures $W^{\text{int}}_{\Uni{7}\times\bbR^+}$ is even smaller 
\begin{alignat}{2}
    W^{\text{int}}_{\SU{7}} &: \repsub{1}{-7}\oplus \repsub{\Bar{7}}{-3}\oplus \repsub{21}{-1}\oplus \repsub{35}{-5}\oplus \text{c.c.}  &\subset \repsub{912}{1} , \\
    W^{\text{int}}_{\Uni{7}\times\bbR^+} &
        : \repsub{1}{-7}\oplus \repsub{35}{-5}\oplus \text{c.c.}
        &\subset \repsub{912}{1} . \label{eq:ECStorsion}
\end{alignat}
In \cite{1910.04795} it was shown that the the conditions of vanishing intrinsic torsion can be split into two simple geometric conditions. The first is the Dorfman involutivity of the $L_3$ bundle,
\begin{equation}\label{def:dorfman-derivative}
    L_V W \in \Gamma(L_3) \text{ for all } V, W \in \Gamma(L_3),
\end{equation}
is necessary and sufficient to set the $\repsub{1}{-7}\oplus \repsub{35}{-5}$ components of the intrinsic torsion to zero. The operator $L_V$ here is the generalised Lie derivative (or Dorfman bracket). For each $V\in\Gamma(E)$ it induces the infinitesimal action of diffeomorphisms and form-field gauge transformations, which we refer to collectively as ``generalised diffeomorphisms'' $\GDiff$. Involutive $L_3$ bundles (or their corresponding $\mathcal{J}$)  are called \textit{integrable} since from~\eqref{eq:ECStorsion} we see that they define a torsion-free $\Uni7\times\bbR^+$ structure. By analogy with the conventional case, we call them \emph{exceptional complex structures}. 

The second set of conditions correspond to the vanishing of a moment map. One
considers a subset of the possible $\SU{7}$-structures for which the corresponding $\UniR{7}$ structure is integrable,
\begin{equation}
    \hat{\mathcal{Z}} := \{\psi\ |\ \mathcal{J}\text{ is integrable } \}.
\end{equation}
It was argued in~\cite{1910.04795} that $\hat{\mathcal{Z}}$ posses a natural psuedo-Kähler metric, inherited from the pointwise action of $\ExR{7(7)}$ on $\psi$. Furthermore, the geometry is invariant under the action of generalised diffeomorphisms on $\psi$ via the Dorfman bracket $L_V$. Since $L_V$ generates this action on the space of $\SU{7}$ structures, we can associate the Lie algebra action of $\gdiff$ with the space of generalised vectors $V\in\Gamma(E)$, and we hence get the corresponding moment map 
\begin{align}
    \mu(V) = -\frac{1}{3}\int_M \big(s(L_V\psi, \bar{\psi})\big)\big(\ii s(\psi, \bar{\psi})\big)^{-2/3}, \quad V\in \Gamma(E)
\end{align}
where the definitions of $s(\psi,\bar{\psi})$ etc are given in~\cite{1910.04795}. The moment map vanishes if and only if the remaining components of $W^{\text{int}}_{SU(7)}$ vanish. Thus a torsion-free $\SU7$ structure is one where the induced $L_3$ is involutive and the moment map vanishes.  

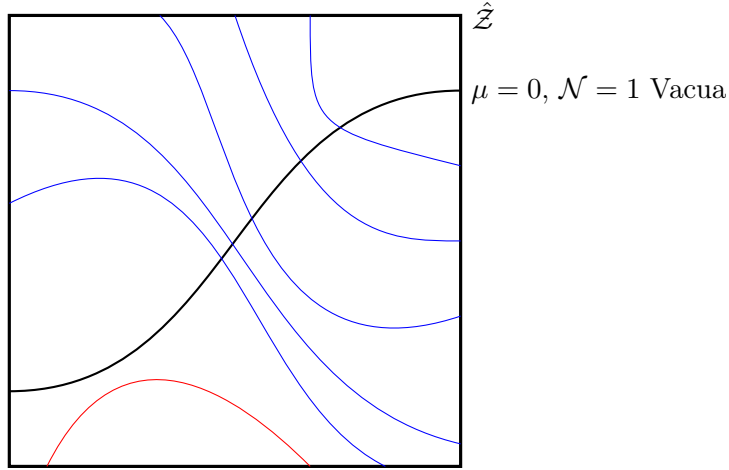
\begin{figure}[h]
\begin{center}
\begin{tikzpicture}
\draw[very thick] (0,0) rectangle (6,6) node[anchor = west] {$\hat{\mathcal{Z}}$};
\draw[black, thick] (0,1) ..controls (3,1) and (3,5)..(6,5) node[anchor = west] {$\mu = 0$,  $\mathcal{N}=1$ Vacua};
\draw[blue] (2,6) .. controls (3,5) and (3,1)..(6,2);
\draw[blue] (3,6) .. controls (4,3) and (5,3)..(6,3);
\draw[blue] (0,5) .. controls (3,5) and (3,1)..(6,0.3);
\draw[blue] (4,6) .. controls (4,4.5) .. (6,4);
\draw[blue] (0,3.5) .. controls (3,5) and (3,1)..(5,0);
\draw[red] (0.5,0) .. controls (1.5,2) and (3,1)..(4,0);
\end{tikzpicture}
\end{center}
\caption{The space of $SU(7)$ structures with involutive $L_3$ bundles, $\hat{\mathcal{Z}}$.}
\label{fig:orbits}
\end{figure}

There is a useful way of viewing the moment map condition that comes from Geometrical Invariant Theory (GIT). Given a Kähler space such as $\hat{\mathcal{Z}}$, satisfying $\mu=0$ is equivalent to being on a ``stable orbit''. The idea is to consider the action of the complexified group\footnote{Strictly in this case, there is no notion of a complexified group, but one can still complexify the action of $L_V$ to give complexified orbits.} $\GDiff_\bbC$ on $\hat{\mathcal{Z}}$. These complex orbits are shown heuristically in Figure~\ref{fig:orbits}.  Orbits which intersect the $\mu=0$ subsapce are called stable and are shown in blue, while orbits which do not intersect are called ``unstable", shown in red. The problem of solving $\mu=0$ then becomes equivalent to determining which orbits are stable, a classic notion in GIT.
Crucially this also gives an alternative picture for counting moduli. One natural approach is to deform the structure, impose vanishing intrinsic torsion (that is the  involutivity and moment map conditions) and then quotient by $\GDiff$, to count the non-equivalent deformations. However, under the complexified orbit GIT picture, one can also impose just involutivity and then quotient by $\GDiff_\bbC$. In this second framework everything depends only on the exceptional complex structure. 

In the following, we discuss the exceptional complex structure involutivity conditions in the case of type-0 and type-3 structures. The former case was essentially considered in~\cite{1910.04795} though here we will discuss in more detail the role of flux sources. The type-3 discussion is new. However, before proceeding we make an important observation. Thus far, there was no reason to exclude the possibility of the structure type changing as one moved in $M$. However, imposing the torsion-free conditions we immediately see from the bilinear equation~\eqref{eq:Lukas-and-saffin-differential-equation-a} that
\begin{equation}
    \dd(\ee^{3\Delta}\sin\chi) = 0 
\end{equation}
or equivalently $\sin\chi=\text{const.}\times\ee^{-3\Delta}$. Thus we cannot have changing from type-0 ($\sin\chi\neq 0$) to type-3 ($\sin\chi=0$) without the warp-factor $\ee^{2\Delta}$ in the metric~\eqref{def:warped-metric} diverging and the eleven-dimensional space becoming singular. Thus we find
\begin{equation}
    \text{for smooth integrable $\SU7$ structures the type cannot change over $M$.}
\end{equation}
As we have noted in the introduction, for generic compact backgrounds we need sources for the fluxes at which points the geometry may be singular. Thus the lack of type-changing should be interpreted as a condition away from sources. 

\subsection{Integrable type-0 ECS}
Given the form~\eqref{def:type_0_original}, a generic element $V\in\Gamma(L_{3})$ can be written as $V=\ee^{\alpha+\beta}\cdot v$ where $v\in\Gamma(T_\bbC)$ and
\begin{align*}
    \alpha = A+\tilXi{3}/\tilXi{}, \qqq \beta=\tilde{A}-\frac{1}{2}A\wedge (\tilXi{3}/\tilXi{}).
\end{align*}
The involutivity conditions for a generic type-0 structure are~\cite{1910.04795}
\begin{align*}
    L_{V}W = L_{\ee^{\alpha+\beta}\cdot v} \ee^{\alpha+\beta}\cdot w = \ee^{\alpha+\beta}\cdot L^{\dd\alpha + \dd\beta -\frac{1}{2}\alpha\wedge \dd\alpha}_{v} w \in \Gamma (L_3) \qquad 
    \forall v,w \in \Gamma(T_\bbC) . 
\end{align*}
where $L^{\Sigma}_V$ is the ``twisted generalised Lie derivative'' defined, for example, in~\cite{1910.04795}. This simplifies to the requirement 
\begin{align}
    L^{\dd\alpha + \dd\beta -\frac{1}{2}\alpha\wedge \dd\alpha}_{v} w = [v,w] + i_w i_v \dd\alpha + i_w i_v (\dd\beta -\tfrac{1}{2}\alpha\wedge \dd\alpha)\in \Gamma(T_\bbC)
\end{align}
for any vector fields $v,w$, and hence
\begin{equation}
    \dd\alpha =0,\qquad \dd\beta = 0 . 
\end{equation}
Given the bilinear expressions~\eqref{eq:ab-blinear-0} the first condition becomes
\begin{align}
    \dd\big(\ee^{2\Delta}\tilXi{(3)}\big) - \big(\ee^{2\Delta}\tilXi{}\big)^{-1}\dd\big(\ee^{2\Delta}\tilXi{}\big)\wedge \big(\ee^{2\Delta}\tilXi{(3)}\big) &=  - \ee^{2\Delta}\tilXi{}F ,
    \label{eq:Invol_four_form_condition}
\end{align}
which we see is a combination of equations~\eqref{eq:Lukas-and-saffin-differential-equation-a} and~\eqref{eq:Lukas-and-saffin-differential-equation-d}. In terms of the $\SU3$ structure it reads
\begin{equation}
\label{eq:type0-F-SU3}
    F = - \dd \left(\cot\chi\, \Omega_+\right) , \qquad
    \dd\left( \csc\chi\, \Omega_- - \sigma\wedge \omega\right) = 0 . 
\end{equation}
Substituting the first condition into the second gives
\begin{equation}
\label{eq:type0-tildeF}
    \tilde{\Xi} \tilde{F} = - \tfrac12 F \wedge \tilXi{(3)} ,    
\end{equation}
reproducing~\eqref{eq:Lukas-and-saffin-differential-equation2a}.  One can relate the involutivity conditions to the extremisation of the superpotential. From the general formula given in ~\cite{1910.04795}, one finds  
\begin{align}
    \mathcal{W} &= \int_M \ee^{2\Delta}\tilde{\Xi}\left(\dd\beta  - \tfrac{1}{2}\dd\alpha\wedge\alpha\right) \\&= \int_M\ee^{2\Delta}\tilde{\Xi}\left(\tilde{F} -\ii F\wedge (\tilXi{3}/\tilXi{}) - \tfrac{1}{2}(\tilXi{3}/\tilXi{})\wedge \dd(\tilXi{3}/\tilXi{})\right)
\end{align}
Setting $\mathcal{W}=\delta\mathcal{W}/\delta\psi=0$ we reproduce $\dd\alpha=0$ and $\dd\beta=0$ from the $\alpha$ and $\tilXi{}$ variations respectively. However, from the $\beta$ variation we also get the extra differential condition~\eqref{eq:Lukas-and-saffin-differential-equation-a}:
\begin{equation}\label{eq:no-type-change}
    \dd(\ee^{2\Delta}\tilXi{}) =0 , 
\end{equation}
so that together with~\eqref{eq:Invol_four_form_condition} and~\eqref{eq:type0-tildeF} we reproduce all the equations given by Lukas and Saffin that are linear in $\tilXi{3}$ and $\tilXi{}$.

\subsection{Integrable type-3 ECS}

Recall from~\eqref{eq:gen-type3-L} that the type-3 bundle can be written as an $\Ex{7,\bbC}$ twist of the bundle $T^{[0,1]} \oplus \Lambda^{[2,0]}$.  The involutivity condition for type-3 reduces to
\begin{align}
    L_{v+\lambda}^{\dd\alpha + \dd\beta - \frac{1}{2}\alpha\wedge \dd\alpha} (w + \eta) \in \Gamma\big(T^{[0,1]}\oplus \Lambda^{[2,0]}\big)
    \qquad \forall v,w\in \Gamma(T^{[0,1]}), \ \lambda,\eta\in \Gamma(\Lambda^{[2,0]}), 
\end{align}
where $\alpha= A - i\sigma\wedge\omega$ and $\beta = \Tilde{A}-\tfrac{1}{8}\Omega\wedge\bar{\Omega}+\tfrac{1}{2}\ii A\wedge\sigma\wedge\omega$.  This condition produces richer constraints than the type-0 case, as terms in $\dd\alpha$ and $\dd\beta-\tfrac{1}{2}a\wedge \dd\alpha$ which mix vectors and $2$-forms do not necessarily vanish. Involutivity implies the following 
\begin{align}
    [v,w] &\in \Gamma(T^{[0,1]}) , \\
    \mathcal{L}_v \eta - \imath_w \dd\lambda + \imath_w\imath_v \dd\alpha &\in \Gamma(\Lambda^{[2,0]}) , \label{eq:type-3-inv-b} \\
    \imath_w\imath_v\left(\dd\beta - \tfrac12\alpha\wedge\dd\alpha \right) &= 0  \label{eq:type-3-inv-c}. 
\end{align}
These translate into 
\begin{itemize}
    \item The distribution $T^{[0,1]}\subset T_\bbC$ is involutive: $[T^{[0,1]}, T^{[0,1]}] \subset T^{[0,1]}$. If $\gamma\in\Gamma(\Lambda^{[p,0]})$, this implies 
    \begin{align}
    \label{eq:T01-lemma}
        0 = \imath_{[v,w]} \gamma = \mathcal{L}_v (\imath_w\gamma) - \imath_w (\mathcal{L}_v\gamma) = - \imath_w (\mathcal{L}_v\gamma) = -\imath_w(\imath_v \dd \gamma) , 
    \end{align}
    In terms of $\Omega$, one can rewrite the involutivity of $T^{[0,1]}$ as $\imath_{[v,w]}\Omega = 0$ if and only if $\imath_v\Omega = 0$ and $\imath_w\Omega = 0$. Taking $\gamma$ to be $\Omega$ in~\eqref{eq:T01-lemma}, we see that this is equivalent to $\imath_u\imath_v\dd\Omega=0$, that is $\dd\Omega\in \Gamma(\Lambda^{[3,1]})$, or  
    \begin{equation*}
        \dd\Omega = X \wedge\Omega, \qquad \qquad  X\in\Gamma\big(\Lambda^{[0,1]}\big) ,
    \end{equation*}
    which reproduces \eqref{eq:Lukas-and-saffin-differential-equation-d} up to the fact that the one-form field $X$ is proportional to $\dd\Delta$.
    \item  Taking $\gamma$ to be $\eta$ or $\lambda$ in~\eqref{eq:T01-lemma}, we see that $\mathcal{L}_v \eta - \imath_w \dd\lambda \in \Gamma(\Lambda^{[2,0]})$. The condition~\eqref{eq:type-3-inv-b} then becomes 
    \begin{equation}
    \label{eq:type-0-F-cond}
        F - \ii \dd (\sigma\wedge \omega) \in \Gamma(\Lambda^{[2,2]}),
    \end{equation}
    \item Finally~\eqref{eq:type-3-inv-c} implies $\dd\beta -\frac12\alpha\wedge\dd\alpha =0$ or equivalently 
    \begin{align}
        \tilde{F} + \tfrac12 (\sigma\wedge\omega)\wedge \dd (\sigma\wedge \omega) &= 0 , \\
    F\wedge\sigma\wedge\omega +\frac{1}{8}\ii \dd (\Omega \wedge \bar{\Omega} )  &= 0. \label{eq:vol-cond}
    \end{align}
\end{itemize}
The involutivity of $T^{[0,1]}$ implies that one can decompose the exterior derivative into differentials associated to the $\ker\Omega$ distribution
\begin{equation}
    \dd = \del' + \delb' ,
\end{equation}
where $\delb':\Lambda^{[p,q]}\to\Lambda^{[p,q+1]}$ and $\del':\Lambda^{[p,q]}\to\Lambda^{[p+1,q]}$ and $\del'^2=\delb'^2=\del'\delb'+\delb'\del'=0$. Since $\sigma\wedge\omega\in\Gamma(\Lambda^{[1,2]})$, we can then decompose~\eqref{eq:type-0-F-cond} as 
\begin{align}
    F_{[3,1]} &= 0 , &
    F_{[1,3]} - \ii \delb' (\sigma\wedge\omega) &= 0 , &
    F_{[0,4]} &= 0 . 
\end{align}
Again, one can show that the involutivity conditions reproduce a particular mix of the conditions implied by Lukas and Saffin's equations \eqref{eq:Lukas-and-saffin-differential-equations} and~\eqref{eq:Lukas-and-saffin-differential-equations2}. The rest of the conditions are then implied by the vanishing of the moment map. 

Again one can view the involutivity conditions as arising from the extremisation of the superpotential.  The type-3 background arises as the $\chi\to 0$ limit of the type-0 background. Taking this limit in the type-0 superpotential we get
\begin{equation}
\begin{split}
    \mathcal{W} 
    &=\int_M\ee^{2\Delta}\big(\ii\dd(\sigma\wedge\omega)-F\big)\wedge\Omega \\ & \quad
    + \int_M\ee^{2\Delta}\Big(\tfrac{1}{4}\dd\chi\wedge\Omega\wedge\bar{\Omega}
    +\chi\big(\tilde{F} +\tfrac{1}{2}\sigma\wedge\omega\wedge\dd(\sigma\wedge\omega)+\ii F\wedge\sigma\wedge\omega + \tfrac{1}{8}\dd\Omega\wedge\bar{\Omega}\big) \Big) 
    \\&\quad +\mathcal{O}(\chi^2).
\end{split}
\end{equation}
As expected, the leading term agrees with the expressions given in \cite{hep-th/0311119}. Furthermore, we see that the $\Omega$ and $\Delta$ variations reproduce~\eqref{eq:type-0-F-cond}. From the linear term, we see that varying $\chi$ (that is varying away from a type-3 structure)  gives the condition~\eqref{eq:vol-cond}. The $A$ and $\sigma\wedge\omega$ variations on the other hand give $\dd(\ee^{2\Delta}\Omega)=0$, which implies $T^{[0,1]}$ is involutive. As before this last condition is actually slightly stronger than involutivity, constraining part of the remaining intrinsic torsion to be related to $\dd\Delta$. 

\subsection{Fluxes, no-go theorems and sources}

So far in our discussion we have ignored sources for the flux $F$. There is a famous no-go theorem for M-theory compactifications to Minkowski space~\cite{hep-th/0007018}: if the warp factor $\ee^{2\Delta}$ is bounded above and below and $M$ is compact then there are no solutions with non-zero $F$ unless there are negative tension sources.

For supersymmetric Minkowski backgrounds of the type here, preserving supersymmetry and satisfying the Bianchi identities for $F$ imply the equations of motion~\cite{Gauntlett:2002fz}.  Assuming that supersymmetry holds, the sources which allow us to evade the no-go theorem must appear in the Bianchi identity. In particular, we require magnetic sources $\rho_M$ for $F$ so that 
\begin{align}
\label{eq:source}
    \dd F = \rho_M , 
\end{align}
where $\rho_M$ is a five-form. For a completely localised (unsmeared) source, such as an M5-brane, the source takes the form $\rho_M=Q\delta^{(5)}(C_2)$ where $C_2$ is a two-cycle on $M$ on which the fivebrane is wrapped. The total source $\rho_M$ must be trivial in de Rham cohomology in order to satisfy~\eqref{eq:source}, so we must have both negative and postive tension objects.

The local supersymmetry conditions on the $\SU7$ structure follow from the Killing spinor equations without making any assumption on the closure of $F$. However, as we have seen, for a type-0 structure supersymmetry implies 
\begin{equation}
   F = - \dd (\cot\chi \Omega_+) ,
\end{equation}
and hence that $\dd F=0$ wherever $F$ is well defined.  For an isolated source of the form $\rho_M=Q\delta^{(5)}(C_2)$ we find that for $\Sigma_5$ a surface transversely intersecting $C_2$ with a type-0 structure everywhere on its boundary
\begin{equation}
  Q = \int_{\Sigma_5} \rho_M = \int_{\Sigma_5}\dd F = \int_{\del \Sigma_5} F=-\int_{\del\Sigma_5}\dd(\cot\chi\Omega_+) 
  = 0 .
\end{equation}
Even if $F$ is not well defined everywhere in $\Sigma_5$, so long as it is defined and the structure is type-0 on $\del\Sigma_5$ then $Q=0$ and so we cannot have isolated sources. This argument extends to smeared sources (with both positive and negative charges) linking $\del\Sigma_5$, given $F$ is everywhere well defined on $\del\Sigma_5$ regardless of how it links with the source.  This means that any sources capable of being linked with a four dimensional surface must have vanishing net charge, and so must be ``dipole'' in nature. This is not the case for, for example, standard M5-brane or orbifold sources, hence we can conclude that no type-0 backgrounds with conventional sources are allowed. The other possible way of evading the no-go theorem is for the warp factor to diverge and our analysis breaks down, specifically in such a way that we go from a type-0 to type-3 structure on the singular locus; the solution in section~4.2 of~\cite{Lukas:2004ip} is of this type. 

Type-3 backgrounds, on the other hand, may admit non-trivial fluxes and so can have sources. In fact, we will see a completely explicit set-up of this when we discuss heterotic M-theory. 

In summary, we find that if we restrict to ``reasonable'' sources, meaning localised with some net charge and no warp-factor singularities, then the only allowed non-trivial flux supersymmetric backgrounds are type-3.  This conclusion relies only the involutive structure, independent of the exact form of the moment map condition.

\section{Deformations}\label{sec:Deformations}

The type-0 infinitesimal moduli were discussed in~\cite{1910.04795} with $\G_2$ structures as the special case. Let us start by slightly generalising the procedure used there to give expressions that work for both type-0 and type-3 structures. Given the coset structure~\eqref{eq:cosets}, any two ECS are related by an $\Ex{7(7)}$ transformation. Thus, we can write a deformation of $L_3$ as 
\begin{equation}
    L_3' = \ee^{\epsilon M}\cdot L_3 = L_3 + \epsilon M\cdot L_3 + \dots ,
\end{equation}
for some adjoint Lie algebrra element $M\in\Gamma(\ad\tilde{F})$. Given $L_3$ is involutive, the involutivity condition for $L'_3$, to leading order in $\epsilon$, then reads
\begin{equation}
\label{eq:gen-def-cond}
    L_{M\cdot U} V + L_U (M\cdot V) - M\cdot (L_U V) \in \Gamma(L_3) , \qquad
    \forall U,V \in \Gamma(L_3). 
\end{equation}
To satisfy the moment map and identify deformations that simply corresponds to diffeomorphsims and gauge transformations, we need to mod out by complex generalised diffeomorphisms $\GDiff_\bbC$. Thus we need to solve~\eqref{eq:gen-def-cond} up to an equivalence if 
\begin{equation}
   M\cdot L_3 \sim (M + L_X) L_3 
\end{equation}
for some $X\in\Gamma(E_\bbC)$. For both type-0 and type-3 structures we had
\begin{equation}
    L_3 = \ee^{a+b} \Delta , \qquad
    \Delta = \begin{cases} 
       T_\bbC & \text{type-0} \\ 
       T^{[0,1]}\oplus\Lambda^{[2,0]} & \text{type-3}
       \end{cases}
\end{equation}
Defining  $M=\ee^{a+b}\cdot  M_0 \cdot \ee^{-a-b}$ we can then write the deformation condition~\eqref{eq:gen-def-cond} as 
\begin{equation}
\label{eq:gen-def-cond-twist}
    L^\CF_{M_0\cdot U} V + L^\CF_U (M_0\cdot V) - M_0\cdot (L^\CF_U V) \in \Gamma(\Delta) , \qquad
    \forall U,V \in \Gamma(\Delta) , 
\end{equation}
and the equivalence condition as
\begin{equation}
\label{eq:gen-def-equiv-twist}
    M_0\cdot \Delta \sim  (M_0 +  L^\CF_X) \Delta , \qquad X\in \Gamma(E_\bbC) , 
\end{equation}
where locally $\CF=\dd a + \dd b -\frac12 a\wedge\dd a$, and we are using the isomorphism $\ee^{-a-b}\cdot E_\bbC\simeq E_\bbC$. In each case, we will see that the moduli are then counted by a cohomology associated to $\Delta$. For type-0, this is the de Rham cohomology, while for type-3 it is one associated to $\delb'$. 

\subsection{Type-0 moduli}

For type-0 structures we have $\CF=\dd a = \dd b -\frac12 a\wedge\dd a=0$. Furthermore we can parameterise the generic deformations of $\Delta=T_\bbC$ by complex three- and six-forms
\begin{equation}
    M_0 = \alpha + \beta , \qquad 
    \alpha \in\Gamma(\Lambda^3T^*_\bbC), \quad \beta \in \Gamma(\Lambda^6T^*_\bbC) .
\end{equation}
A short calculation shows that the involutivity condition~\eqref{eq:gen-def-cond-twist} reduces to 
\begin{equation}
    \imath_x\imath_y (\dd\alpha + \dd\beta) = 0 , \qquad \forall x,y\in\Gamma(T_\bbC) , 
\end{equation}
or equivalently 
\begin{equation}
    \dd\alpha = 0, \qqq \dd\beta = 0.
\end{equation}
Writing $X=x+\nu+\rho+\tau$ for a generalised vector in $E_\bbC$, the components ordered as in~\eqref{def:generalised_vector}, and given $\mathcal{L}_x T_\bbC=0$, in the sense that the Lie bracket with $x$ does not change the bundle $T_\bbC$ within $E_\bbC$, in the equivalence condition~\eqref{eq:gen-def-equiv-twist} we have  
\begin{equation}
     L^\CF_X T_\bbC = -(\dd\nu + \dd\rho)\cdot T_\bbC
\end{equation}
and so we have 
\begin{equation}
    \alpha \sim \alpha - \dd\nu , \qquad
    \beta \sim \beta - \dd\rho , 
\end{equation}
We see that inequivalent deformations are counted by the third and sixth complex-valued de Rham cohomologies on $M$~\cite{1910.04795}
\begin{align}
    \text{moduli space} = H_\dd^3(M, \bbC) \oplus H_\dd^6(M, \bbC).
\end{align}
In the special case where $M$ is a $\G_2$ holonomy manifold, $H^6_\dd(M, \bbC)$ vanishes and we recover the known result for infinitesimal moduli for closed $\Gx{2}$ structures. In the context of compact $M$, we have already argued that it is not possible to support the flux in type-0 backgrounds with conventional sources, and so, in fact, $\G_2$ holonomy backgrounds are then the only solutions. 

\subsection{Type-3 moduli}\label{subsec:type-3 deformatins}

%
%
For the involutive deformation of a type-3 structure, the generic deformations of $\Delta=T^{[0,1]}\oplus \Lambda^{[2,0]}$ are naturally parameterised by
\begin{equation}\label{def:deformation_section}
    M_0 = \kappa+r+\theta+\mu
\end{equation}
with each component being a section of the complex bundles
\begin{equation}
\label{eq:type3-mod-fields}
\begin{aligned}
    \kappa &\in \Gamma\big(\Lambda^3 T^{[1,0]}\big), &
    r &\in \Gamma\big(T^{[1,0]}\otimes \Lambda^{[0,1]}\big),&
    \theta &\in \Gamma\big(\Lambda^{[0,3]}\oplus \Lambda^{[1,2]}\big),&
    \mu &\in \Gamma\big(\Lambda^6 T_\bbC^*\big).
\end{aligned}
\end{equation}
In addition, $\CF$ no longer vanishes on involutive backgrounds, but rather 
\begin{equation}
    \CF = \dd a = F - \ii \dd (\sigma\wedge\omega) \in \Gamma(\Lambda^{[2,2]}) . 
\end{equation}
The vanishing of the other components and the Bianchi identity for $F$ imply that $\del'\CF=\delb'\CF=0$.  The involutivity constraint~\eqref{eq:gen-def-cond-twist} reduces to the conditions 
\begin{align}
    \delb'\kappa &= 0,\label{eq:chi_diff_eq} \\
    \delb'r + \kappa\cdot \CF &= 0 , \label{eq:r_diff_eq} \\
    \delb'\theta_{[0,3]} &= 0 ,\label{eq:theta_diff_eq_1} \\
    \bar{\del}'\theta_{[1,2]} + \del'\theta_{[0,3]} - r\cdot \CF &= 0 , \label{eq:theta_diff_eq_2} \\
    \delb'\mu_{[3,3]} + \del' \mu_{[2,4]} -  \theta_{[1,2]} \wedge\CF &= 0, \label{eq:mu_diff_eq}
\end{align}
where the contractions with $\CF$ are defined as 
\begin{equation}
\label{eq:contraction-action}
    (\kappa\cdot\CF)^m{}_{np} = \tfrac12 \kappa^{mrs} \CF_{rsnp} , \qquad
    (r\cdot\CF)_{mnpq} = 4 r^s{}_{[m} \CF_{npq]s} . 
\end{equation}
Again writing $X = x + \nu + \rho + \tau$ for a generalised vector, since $L^\CF_{x^{[0,1]}}\Delta=0$ by definition, in the equivalence condition~\eqref{eq:gen-def-equiv-twist} we have
\begin{equation}\label{eq:type-3-gdiff-action}
\begin{split}
     L^\CF_X \Delta 
    = -(\bar{\del}' x^{[1,0]}
    + \dd\nu
    -\imath_{x^{[1,0]}}\CF + \dd\rho-\CF\wedge\nu) \cdot \Delta.
\end{split}
\end{equation}
We therefore get the following equivalences
\begin{equation}
\begin{aligned}
    r &\sim r-\bar{\del}' x^{[1,0]}, \\
    \theta_{[0,3]} &\sim \theta_{[0,3]}-\bar{\del}'\nu_{[0,2]}, \\ 
    \theta_{[1,2]} &\sim \theta_{[1,2]}-(\del'\nu_{[0,2]}+\bar{\del}'\nu_{[1,1]}-\imath_{x^{[1,0]}}\CF), \\
    \mu_{[2,4]} &\sim \mu_{[2,4]} - (\delb'\rho_{[2,3]} + \del'\rho_{[1,4]}-\nu_{[0,2]}\wedge\CF) \\
    \mu_{[3,3]} &\sim \mu_{[3,3]} - (\delb'\rho_{[3,2]} + \del'\rho_{[2,3]}-\nu_{[1,1]}\wedge\CF) 
\end{aligned}
\label{eq:type-3-gdiff-action}
\end{equation}
Equations \eqref{eq:chi_diff_eq}--\eqref{eq:mu_diff_eq} and \eqref{eq:type-3-gdiff-action} have a symmetry parameterised by $\gamma\in\Gamma(\Lambda^{[2,1]})$
\begin{equation}
\label{eq:CF-shift}
\begin{gathered}
    \CF \mapsto \CF + \delb'\gamma , \\
    r \mapsto r - \kappa\cdot\gamma , \qquad
    \theta \mapsto  \theta + r\cdot\gamma , \qquad
    \mu \mapsto \mu - \gamma \wedge \theta , \\
    \nu \mapsto \nu + \imath_{x^{[1,0]}} \gamma , \qquad
    \rho \mapsto \rho + \nu \wedge \gamma . 
\end{gathered}
\end{equation}
This implies that the deformation problem only depends on the cohomology group defined by $\delb'$ 
\begin{equation}
    [ \CF ] \in H^{[2,2]}_{\delb'}(M) . 
\end{equation}
If we have a Hodge decomposition so that we can decompose de Rham cohomologies into $\delb'$ cohomologies, since $\CF$ and $F$ differ by a $\dd$-exact pice, it is precisely the physical flux $[F]\in H^{[2,2]}_{\delb'}(M)$ that enters the problem. This will be the case for the heterotic M-theory example we consider in section~\ref{sec:Heterotic M-Theory}. For now, we will not make this assumption and keep $\CF$. 

We also see that in general, the flux can obstruct moduli. For example, for $\kappa$, from~\eqref{eq:chi_diff_eq} it would appear that $\kappa\in H^0_{\delb'}(\ext^3T^{[1,0]})$. However~\eqref{eq:r_diff_eq} implies that $\kappa\cdot\CF$ is $\delb'$-exact. If $\CF$ is non-trivial in $ H^{[2,2]}_{\delb'}(M)$ this is not possible, unless $\kappa=0$. Thus the non-triviality of $\CF$ obstructs the naïve modulus in $\kappa$.  Note that $\kappa$ deforms the type-3 structure into a type-0 structure, as may be seen by its action on the type-3 $\SU{7}$ tensor $\psi$, contracting with the leading three form to produce a scalar. Thus the obstruction of the $\kappa$ deformation is consistent with supersymmetry prohibiting local type change in the presence of flux.


To analyse the deformation problem further, we can make the assumption that we have a $\del'\delb'$-lemma, namely
\begin{equation}
\label{eq:ddbar-prime}
    \image\del' \cap \ker \delb' = \image\del' \cap \ker \delb' = \image\del'\delb' . 
\end{equation}
A sufficient condition for this to hold is that $M$ admits a metric of the form~\eqref{eq:metric-SU3} where the $\SU3$ metric is Kähler and has the same $\Omega$ structure as defines the $T^{[0,1]}$ bundle and $\dd\sigma=0$. 
Since $\delb'\theta_{[0,3]}=0$, the lemma implies that we can write
\begin{equation}
    \del'\theta_{[0,3]} = \delb'\del'\zeta_{[0,2]}, \qquad
    \del'\mu_{[2,4]} + \del'\zeta_{[0,2]}\wedge\CF = \delb'\del'\xi_{[1,3]} , 
\end{equation}
for some $\zeta_{[0,2]}$ and $\xi_{[1,3]}$, where in the second expression we have used the fact that $\delb'(\del'\mu_{[2,4]} + \del'\zeta_{[0,2]}\wedge\CF)=0$ identically and $\del'\zeta_{[0,2]}\wedge\CF=\del'(\zeta_{[0,2]}\wedge\CF)$, and hence the lemma can again be applied. We define
\begin{equation}
    \theta'_{[1,2]} = \theta_{[1,2]} + \del'\zeta_{[0,2]} , \qquad
    \mu'_{[3,3]} = \mu_{[3,3]} + \del'\xi_{[1,3]} , 
\end{equation}
so that the conditions \eqref{eq:chi_diff_eq}--\eqref{eq:mu_diff_eq} become
\begin{equation}
\label{eq:simpler-closed}
    \begin{aligned}
    \delb'\kappa &= 0, & && && 
    \delb'r + \kappa\cdot \CF &= 0 , \\
    \delb'\theta_{[0,3]} &= 0 , & && &&
    \bar{\del}'\theta'_{[1,2]}  - r\cdot \CF &= 0 , \\
    \delb'\mu'_{[3,3]} -  \theta'_{[1,2]} \wedge\CF &= 0. 
\end{aligned}
\end{equation}
By similar use of the lemma the equivalence relations become
\begin{equation}
\label{eq:simpler-equiv}
\begin{aligned}
    r &\sim r-\bar{\del}' x^{[1,0]}, \\
    \theta_{[0,3]} &\sim \theta_{[0,3]}-\bar{\del}'\nu'_{[0,2]}, \\ 
    \theta'_{[1,2]} &\sim \theta'_{[1,2]}-(\bar{\del}'\nu_{[1,1]}-\imath_{x^{[1,0]}}\CF), \\
    \mu'_{[3,3]} &\sim \mu'_{[3,3]} - (\delb'\rho'_{[3,2]} - \nu'_{[1,1]}\wedge\CF) 
\end{aligned}
\end{equation}
We can now calculate the corresponding moduli space in terms of cohomology groups. We first note that if $[\CF]$ is trivial, we can use the symmetry~\eqref{eq:CF-shift} to set $\CF$ to zero and we get the moduli space 
\begin{equation}
\begin{aligned}
    \text{moduli space for ${\CF}$ trivial} &= H^0_{\delb'}(\ext^3T^{[1,0]})  
        \oplus H^{1}_{\delb'}(T^{[1,0]})
        \\ & \qquad \qquad 
       \oplus  H^{[0,3]}_{\delb'}(M) \oplus  H^{[1,2]}_{\delb'}(M) \oplus  H^{[3,3]}_{\delb'}(M) . 
\end{aligned}
\end{equation}
If we  turn to the case of non-trivial $[\CF]$, 
we see that the $\theta_{[0,3]}$ moduli are still counted by $H^{[0,3]}_{\delb'}(M)$, that is 
\begin{equation}
    \text{$\theta_{[0,3]}$ moduli space} = H^{[0,3]}_{\delb'}(M) . 
\end{equation}
To calculate the moduli space of the remaining deformations we start by noting that the operators acting on the moduli fields in~\eqref{eq:simpler-closed} and~\eqref{eq:simpler-equiv} can be viewed as the total differential for a double complex. Consider figure \ref{fig:moduli_double_complex},
\begin{figure}
    \centering
    \begin{equation}
    \begin{tikzcd}
    & & & \vdots \arrow[d] & \vdots \arrow[d,"\delb'"] & \\
    & & 0 \arrow[r] &
        \Lambda^{[1,0]} \arrow[r,"\wedge\CF"] \arrow[d,"\delb'"] &
        \Lambda^{[3,2]} \arrow[r] \arrow[d,"\delb'"] & 0 \\
    & & 0 \arrow[d] \arrow[r] &
        \Lambda^{[1,1]} \arrow[r,"\wedge\CF"] \arrow[d,"\delb'"] &
        \Lambda^{[3,3]} \arrow[r] \arrow[d,"\delb'"] & 0 \\
    &0\arrow[r] & T^{[1,0]}\otimes \Lambda^{[0,0]} 
        \arrow[r,"\cdot\CF"] \arrow[d,"\delb'"] &
        \Lambda^{[1,2]} \arrow[r,"\wedge\CF"] \arrow[d,"\delb'"] &
        \Lambda^{[3,4]} \arrow[r] \arrow[d] & 0 \\
    &0\arrow[d] \arrow[r] & T^{[1,0]}\otimes \Lambda^{[0,1]} 
        \arrow[r,"\cdot\CF"] \arrow[d,"\delb'"] & 
        \Lambda^{[1,3]} \arrow[r] \arrow[d,"\delb'"] & 0 & \\
    0\arrow[r]&\Lambda^3T^{[1,0]}\otimes\Lambda^{[0,0]}\arrow[d,"\delb'"] \arrow[r, "\cdot \CF" ] & T^{[1,0]}\otimes \Lambda^{[0,2]} 
        \arrow[r,"\cdot\CF"] \arrow[d,"\delb'"] & 
        \Lambda^{[1,4]} \arrow[r] \arrow[d]  & 0 & \\
    0\arrow[r]& \Lambda^3T^{[1,0]}\otimes\Lambda^{[0,1]} \arrow[d,"\delb'"] \arrow[r, "\cdot \CF" ] & T^{[1,0]}\otimes \Lambda^{[0,3]}\arrow[d,"\delb'"] \arrow[r] & 0& &\\
    &\vdots &\vdots & & &
    \end{tikzcd}
    \end{equation}
    \caption{The double complex used to compute the cohomology groups for the differential graded algebra of deformations, gauge transformations and generalised torsion.}
    \label{fig:moduli_double_complex}
\end{figure}
where the $\cdot\CF$ action on vector valued forms is the action given in~\eqref{eq:contraction-action}.  This defines a double complex as $\CF$ is $\delb'$-closed and $\imath_x\CF\wedge\CF=0$ for all vectors $x$, and similarly for $\kappa$ (that is $(\kappa\cdot\CF) \cdot \CF=0$). The total complex is formed by combining terms on the diagonals to give 
\begin{equation}
\begin{tikzcd}
    \cdots \arrow[r,"\delb'+\CF"] & U_{p-1} \arrow[r,"\delb'+\CF"] & U_p \arrow[r,"\delb'+\CF"] & U_{p+1} \arrow[r,"\delb'+\CF"] & \cdots 
\end{tikzcd}
\end{equation}
where 
\begin{equation}
    U_p = \bigl(\Lambda^3 T^{[1,0]}\otimes \Lambda^{[0,p-3]} \bigr)\oplus \bigl(T^{[1,0]}\otimes\Lambda^{[0,p-2]}\bigr) \oplus \Lambda^{[1,p-1]}  
    \oplus \Lambda^{[3,p]} .
\end{equation}
We see that indeed we get exactly the operators that appear in~\eqref{eq:simpler-closed} and~\eqref{eq:simpler-equiv} so that, in particular, the remaining moduli are counted by the $(\delb'+\CF)$-cohomology group
\begin{equation}
    \text{$(\kappa + r+\theta'_{[1,2]}+\mu'_{[3,3]})$ moduli space} = H^{3}_{\delb'+\CF} , 
\end{equation}
so that we have in general 
\begin{equation}
  \label{eq:type-3-moduli-space}
  \text{type-3 moduli space} = H^{3}_{\delb'+\CF}
       \oplus  H^{[0,3]}_{\delb'}(M) . 
\end{equation}  
One can then use  standard spectral sequence techniques to calculate the cohomology of $\delb'+\CF$ from the double complex. This procedure corresponds to solving the equations~\eqref{eq:simpler-closed} and~\eqref{eq:simpler-equiv} in a sequence of steps. Using the vertical filtration we get the first page of the spectral sequence in figure \ref{fig:spectral_seq_P1},
\begin{figure}
    \centering
    \begin{equation}
    \begin{tikzcd}
    & & & \vdots & \vdots & \\
    &  & 0 \arrow[r] &
        H^{[1,1]}_{\delb'}(M) \arrow[r,"\wedge{[\CF]}"] &
        H^{[3,3]}_{\delb'}(M) \arrow[r]  & 0 \\
    & 0 \arrow[r] & H^0_{\delb'}(T^{[1,0]})
        \arrow[r,"\cdot{[\CF]}"] &
        H^{[1,2]}_{\delb'}(M) \arrow[r,"\wedge{[\CF]}"] &
        H^{[3,4]}_{\delb'}(M) \arrow[r]  & 0 \\
    0\arrow[r,]& H^0_{\delb'}(\Lambda^3 T^{[1,0]}) \arrow[r,"\cdot{[\CF]}"] & H^1_{\delb'}(T^{[1,0]})
        \arrow[r,"\cdot{[\CF]}"] &
        H^{[1,3]}_{\delb'}(M) \arrow[r,] &
        0   &  \\
    & \vdots & \vdots & & 
    \end{tikzcd}
    \end{equation}
    \caption{The first page, with the maps $d_1=[\CF]$ shown}
    \label{fig:spectral_seq_P1}
\end{figure}
where the groups on the first page are the homology groups for the vertical $\delb'$ maps. These groups would give moduli space if the flux were trivial. On this first page we show the $\cdot[\CF]$ and $\wedge[\CF]$ actions on $\delb'$ cohomology classes, which is well defined since $\CF$ is $\delb'$-closed. 

Finding the cohomology of these horizontal maps takes us to the second page shown in figure \ref{fig:spectral_seq_P2}.
\begin{figure}
    \centering
    \begin{equation}
    \begin{tikzcd}
    & & \vdots & \vdots & \\
     & 0  &
        H^{[1,1]}_{\delb',\CF}(M)  &
        H^{[3,3]}_{\delb',\CF}(M) & 0 \\
     0  & H^0_{\delb',\CF}(T^{[1,0]}) \arrow[rru,"\delta"] &
        H^{[1,2]}_{\delb',\CF}(M) &
        H^{[3,4]}_{\delb',\CF}(M) & 0 \\
     0  & H^1_{\delb',\CF}(T^{[1,0]}) \arrow[rru,"\delta'"] &
        H^{[1,3]}_{\delb',\CF}(M) &
        0 &  \\
    & \vdots & \vdots & & 
    \end{tikzcd}
    \end{equation}
    \caption{The second Page with the nontrivial $d_2$ maps shown}
    \label{fig:spectral_seq_P2}
\end{figure}
The abelian groups in the complex are the $[\CF]$-cohomology groups 
\begin{equation}
\begin{aligned}
    H^p_{\delb',\CF}(\Lambda^3T^{[1,0]})  &= 
        \ker \Bigl( H^{p}_{\delb'}(\Lambda^3 T^{[1,0]})\xrightarrow{\cdot[\CF]} H^{p+2}_{\delb'}(T^{[1,0]}) \Bigr) = 0  \\
    H^p_{\delb',\CF}(T^{[1,0]})  &= 
        \ker \Bigl( H^{p}_{\delb'}(T^{[1,0]})\xrightarrow{\cdot[\CF]} H^{[1,p+2]}_{\delb'}(M) \Bigr)  \\
    H^{[1,p]}_{\delb',\CF}(M)  &= 
        \ker \Bigl( H^{[1,p]}_{\delb'}(M)\xrightarrow{\wedge[\CF]} H^{[3,p+2]}_{\delb'}(M) \Bigr) \bigg/
        \image\Bigl( H^{p-2}_{\delb'}(T^{[1,0]})\xrightarrow{\cdot[\CF]} H^{[1,p]}_{\delb'}(M) \Bigr) , \\
    H^{[3,p]}_{\delb',\CF}(M)  &= 
        H^{[3,p]}_{\delb'}(M) \bigg/
        \image\Bigl( H^{[1,p-2]}_{\delb'}(M)\xrightarrow{\wedge[\CF]} H^{[3,p]}_{\delb'}(M) \Bigr) . 
\end{aligned}
\end{equation}
We could predict the appearance of these groups from ~\eqref{eq:simpler-closed}, as from these equations get a set of obstructions from the requirement that the following elements of cohomology are trivial, 
\begin{equation}
\label{eq:obstructions}
    [\kappa \cdot \CF] = 0 , \qquad
    [r \cdot  \CF ] = 0 , \qquad
    [\theta'_{[1,2]} \wedge  \CF ] = 0 . 
\end{equation}
These give the kernels in the $[\CF]$-cohomology. Furthermore we see from~\eqref{eq:simpler-equiv} that having used $x^{[1,0]}$ to gauge $r$, we still have residual gauge transformation with $\delb'x^{[1,0]}=0$. These can be used to remove $[\theta'_{[1,2]}]$ of the form $[\imath_{x^{[1,0]}}\CF]$. A similar discussion holds for $[\nu'_{[1,1]}\wedge \CF]$ and together this gives the modding out of images in the $[\CF]$-cohomology. Since by assumption $[\CF]\neq 0$ we have $H_{\delb', \CF}^p(\Lambda^3T^{[1,0]}) = 0$. This reduces the number of groups and pages in the spectral sequence that need to be considered.  

The only non-trivial maps on the second page are $\delta$ and $\delta'$ as shown, and the third page differs from the second only by the cohomology of these two maps.  In the case that the two maps $\delta$ and $\delta'$ are trivial then the third page would be identical to the second and the spectral sequence would stabilise.  We could then simply use the $H^{\bullet}_{\delb',\CF}$ cohomology groups to compute the cohomology groups for the total complex operator $\delb'+F$ 
\begin{align}
    H^p_{\delb'+\CF}\cong  H^{p-2}_{\delb',\CF}(T^{[1,0]})\oplus H^{[1,p-1]}_{\delb',\CF}\oplus H^{[3,p]}_{\delb',\CF}.
\end{align}
In this case the deformations of the exceptional complex structure are counted by 
\begin{equation}
    H^{[0,3]}_{\delb'}\oplus H^3_{\delb'+\CF}\cong  H^1_{\delb',\CF}(T^{[1,0]})\oplus H^{[1,2]}_{\delb',\CF}\oplus H^{[0,3]}_{\delb',\CF}\oplus H^{[3,3]}_{\delb',\CF}.
\end{equation}
If however the $\delta$ and $\delta'$ maps on the second page are non-trivial then we would have one further page in the spectral sequence (figure \ref{fig:spectral_seq_P3}) before the sequence stabilises.
\begin{figure}
    \centering
    \begin{equation}
    \begin{tikzcd}
    & & \vdots & \vdots & \\
     & 0  &
        H^{[1,1]}_{\delb',\CF}(M) &
        H^{[3,3]}_{\delb',\CF}(M)/\image\delta & 0 \\
     0  & \ker \delta  &
        H^{[1,2]}_{\delb',\CF}(M) &
        H^{[3,4]}_{\delb',\CF}(M)/\image\delta' & 0 \\
     0  & \ker\delta' &
        H^{[1,3]}_{\delb',\CF}(M) &
        0 &  \\
    & \vdots & \vdots & & 
    \end{tikzcd}
    \end{equation}
    \caption{The third page, $d_3=0$}
    \label{fig:spectral_seq_P3}
\end{figure}
The third page measures the amount by which the cohomology groups of the total complex are smaller than the sum of the corresponding $[\CF]$-cohomology groups and the deformations of the exceptional complex structure involutive deformations are now counted by 
\begin{equation}\label{eq:Third_page_cohomology}
     H^{[0,3]}_{\delb'}\oplus H^3_{\delb'+\CF}\cong  \ker\delta'\oplus H^{[1,2]}_{\delb',\CF}\oplus H^{[0,3]}_{\delb',\CF} \oplus  H^{[3,3]}_{\delb',\CF}/\image\delta.
\end{equation}
We can understand the $\delta'$ map through the equations \eqref{eq:simpler-closed}.  Let $r$ be a solution to these equations, making it both $\delb'$ and $[\CF]$-closed with 
\begin{align*}
    r\cdot \CF = \delb'\theta_{[1,2]}'.
\end{align*}
where $\theta_{[1,2]}'$ is treated as an defined up to any $\delb'$-closed form $\Delta$ so $ \theta_{[1,2]}=\theta_{[1,2]}'+\Delta$.  Consider then the implicitly $r$-dependent $\delb'$-closed quantity and its equivalence class in $H_{\delb'}^{[3,4]}(M)$
\begin{align}
    \theta_{[1,2]}'\wedge \CF \implies [\theta_{[1,2]}'\wedge \CF]
\end{align}
which is not uniquely defined but is instead defined up to the addition of any element $\Delta\wedge\CF$ with $\delb'\Delta=0$
\begin{align}
    [\theta_{[1,2]}'\wedge \CF]\sim [\theta_{[1,2]}'\wedge \CF]+[\Delta]\wedge[\CF].
\end{align}
The equivalence class $[\theta_{[1,2]}'\wedge \CF]$ defines an element in $H^{[3,4]}_{\delb'}(M)$ up to the ideal in $H^{[3,4]}_{\delb'}(M)$ generated by $[\CF]$.  We may not split $[\theta_{[1,2]}'\wedge \CF]$ into $[\theta_{[1,2]}']\wedge [\CF]$ as $\theta_{[1,2]}'$ is not $\delb'$-closed and so $[\theta_{[1,2]}']$ does not make sense.  The class is then
\begin{align}
    \llbracket\theta_{[1,2]}'\wedge \CF\rrbracket\in H^{[3,4]}_{\delb'}(M)/([\CF]) = H^{[3,4]}_{\delb',\CF}(M)
\end{align}
and through this process we have produced a well defined mapping
\begin{align}
    \delta':H_{\delb',\CF}^1(T^{[1,0]})\to H^{[3,4]}_{\delb',\CF}(M)\quad \text{ by }\quad  r \mapsto \llbracket\theta_{[1,2]}'\wedge \CF\rrbracket
\end{align}
we see that in order for a solution to the $\mu$ equation in \eqref{eq:simpler-closed} to be possible with the $r$ we began with, we require $r\in\ker\delta'$ and then $\mu_{[3,3]}$ is specified up to a $\delb'$-closed term which we cannot access.  If $r\notin \ker\delta'$ then we may not find a $\mu$ to solve the final equation.  We can do the same with $\delta$.  In the above quotients we have fixed the gauge for $\theta'_{[1,2]}$
\begin{align}
    x\cdot \CF = \delb'\nu_{[1,1]}'
\end{align}
and using the same reasoning as above
\begin{align}
    \delta(x) = \llbracket \nu'_{[1,1]}\wedge\CF\rrbracket\in H^{[3,3]}_{\delb',\CF}
\end{align}
The image of $\delta$ captures the parts of the gauge transformations of $\mu_{[3,3]}$ which cannot be captured by $\delb'\rho_{[3,2]}$ nor $[\Delta']\wedge[\CF]$ where $\Delta'$ is the undetermined $\delb'$ closed part of $\nu_{[1,1]}'$.  The quotient by $\image\delta$ removes the gauge transformations which are neither in the image of $\delb'$ nor $[\CF]$.

\section{Heterotic M-Theory}\label{sec:Heterotic M-Theory}
Having given the generic form of the local moduli space, we would like to apply it to a particular case. The prime case with non-trivial flux is heterotic M-theory~\cite{hep-th/9710208,hep-th/9806051}.  %
Ho\v{r}ava and Witten argued that the strongly coupled dual of the heterotic string is M-theory compactified on an interval $S^{1}/\bbZ_{2}$ with particular gauge matter localised to the end hypersurfaces~\cite{hep-th/9510209,hep-th/9603142}. Heterotic M-theory is the further $\mathcal{N}=1$ supersymmetric compactification to Minkowski space on a six-dimensional manifold $X$ that is topologically Calabi--Yau. Thus it is within the class considered here with $M=X\times S^{1}/\bbZ_{2}$ and is, in fact, type-3. To date, the local moduli space for the heterotic M-theory background has been calculated using only the first-order correction to the full supergravity solution due to the flux. The goal of this section is to show how the moduli space can be computed directly using only the cohomology of $X$ and $S^1/\bbZ_2$, and further that, as expected, it agrees with the dual heterotic string result. 

\subsection{Heterotic M-theory geometry}

Let us start by briefly reviewing the heterotic M-theory geometry and showing how it fits into the $\SU7$ generalised geometry classification.

Ho\v{r}ava and Witten formulated the M-theory compactification on $S^1/\bbZ_2$ as an expansion in powers of the eleven-dimensional Planck length cubed $l_p^3$. To leading order, the theory is just the bulk theory of eleven-dimensional supergravity. The metric is even under the $\bbZ_2$ action, while the flux $F$ is odd (so that the dual seven-form $\tilde{F}$ is even). The leading-order supersymmetric heterotic M-theory solution has vanishing flux (because it is odd) and warp factor, and a metric of the form
\begin{equation}
\label{eq:CY-metric}
  \dd s^2(M) = \dd s_{\text{CY}}^2(X) + \dd y^2 , 
\end{equation}
where $\dd s_{\text{CY}}^2(X)$ is a Ricci-flat Calabi--Yau metric on $X$ and $y$ is the coordinate on $S^1/\bbZ_2$. The $\bbZ_2$ quotient acts on the Killing spinors such that the generalised $\SU7$ structure is type-3. The generalised tensor $\psi$ is thus given by~\eqref{eq:psi_from_spinor_bilinears} with $A=\tilde{A}=\Delta=0$ and
\begin{equation}
  \dd \omega = \dd \Omega =0 , \qquad \sigma=\dd y . 
\end{equation}
To next-order in $l_p^3$, one needs to include gauge matter living on the boundaries of the interval, which is required in order to remove the ten-dimensional boundary gravitational anomaly. In particular the boundaries provide delta-function sources for the four-form field strength $F$. Equivalently they impose boundary conditions
\begin{equation}
\label{eq:flux-sources}
  \left.F\right|_{M_{10}^{(1)}} = j_{(1)} , \qquad 
  \left.F\right|_{M_{10}^{(2)}} = -j_{(2)} 
\end{equation}
where $M_{10}^{(i)}$ with $i=1,2$ are the two ten-dimensional boundaries. The heterotic M-theory background is still of type-3 but now with non-trivial flux because of the sources $j_{(i)}$.  We still have $\dd\Omega=0$ and the complex structure on $X$ is unchanged, however now $\omega$ and $\sigma$ are not necessarily closed.

We should thus be able to calculate the moduli of Heterotic M-theory backgrounds using the cohomology for type-3 geometries. The order $l_p^3$ correction of the solution around the zeroth-order Calabi--Yau background due to the non-trivial flux was found by Witten~\cite{Witten:1996mz} and then used to calculate the moduli in the special case of ``standard embedding'' sources (and excluding the complex structure moduli) in~\cite{hep-th/9806051}. Here we will not need to specify the exact supersymmetric background since we can calculate the cohomology groups that appear in~\eqref{eq:type-3-moduli-space} using only the complex structure of the Calabi--Yau and the product structure of $M$. The only subtlety is that extra cohomology groups will appear because of the $S^1/\bbZ_2$ orbifold. They can either be thought of as arising from freedom in imposing boundary conditions on the interval or as encoding extra cohomological information associated to the fixed points of the $\bbZ_{2}$ action. We will also include general the complex structure deformations and allow for general sources. This way we will see that we match the generic heterotic theory result. 

Before turning to the calculation, it is interesting to consider the range of validity of the cohomology calculation. By construction it applies for the full supergravity solution beyond the linearised solution of Witten. The sources $j_i$ scale as $l_p^3$ and so the next-order correction to Witten's solution appears at $l_p^6$.  At this order there are new contributions to the M-theory bulk action coming from, for example, curvature-to-the-fourth terms.
%
%
Such higher order terms come with corrections to the supersymmetry variations, and we will not include them terms in our analysis.  Thus a priori one is not justified in going beyond Witten's linear correction within the context of uncorrected eleven-dimensional supergravity and so there is not a regime where the calculation of the moduli for the full back-reacted geometry is pertinent. Nonetheless the power here is that cohomological techniques give the moduli without even the need to solve even the linear problem. In addition, they will continue to apply even if the moment map ``D-term'' conditions get corrected provided the involutivity ``F-term'' conditions are unchanged. As discussed in~\cite{1910.04795} the structure $\psi$ can effectively be viewed as an infinite set of $\mathcal{N}=1$ chiral fields in four-dimensions and the involutivity conditions come from extremizing a superpotential. Thus assuming some $\mathcal{N}=1$ perturbative non-renormalisation condition for the superpotential would then imply that the cohomology calculation was valid despite corrections to the eleven-dimensional supergravity.

\subsection{Moduli}

We saw above that in general the moduli of a type-3 
structure are given by the cohomology groups on third page of a spectral sequence, summarised by equation \eqref{eq:Third_page_cohomology}
\begin{align*}
H_{\delb'}^{[0,3]}\oplus H_{\delb'+\CF}^{3} & \cong\ker\delta'\oplus H_{\delb',\CF}^{[1,2]}\oplus H_{\delb'}^{[0,3]}\oplus\left(H_{\delb',\CF}^{[3,3]}/\im\delta\right) 
\end{align*}
We thus need to compute this third page for the heterotic M-theory geometry, although in fact we will see that under our assumptions the spectral sequence stabilises on the second page.
We need to include in our analysis the fact that $M$ has boundaries on which there are sources for fields and deformations.  These sources may be treated as boundary conditions on the fields and cohomology classes on the interval, and so we will have to mildly generalise our cohomology groups to accommodate this.  

Recall that the derivation of the infintesimal moduli space \eqref{eq:Third_page_cohomology} used a $\del'\delb'$-lemma \eqref{eq:ddbar-prime}
\begin{equation*}
    \image\del' \cap \ker \delb' = \image\del' \cap \ker \delb' = \image\del'\delb' . 
\end{equation*}
for the exceptional complex structure on $M$. Topologically we have a product $M=X\times S^1/\bbZ_2$ and $X$ is Calabi--Yau, so $M$ admits a product metric of the form~\eqref{eq:CY-metric} where the Calabi--Yau metric is compatible with the complex structure as defined by the type-3 structure. Hence $X$ satisfies a conventional $\del\delb$-lemma and it follows that~\eqref{eq:ddbar-prime} is satisfied. The Calabi--Yau product metric has an integrable $\SU3\subset\GL{7,\bbR}$ that is is compatible with the integrable structure on $T^{[0,1]}$, namely $\SU3\subset\GL{3, \bbC}\times \bbR \ltimes \bbC^3\subset\GL{7,\bbR}$.  By considering harmonic representatives we can relate $\delb'$ on $M$, to de Rham and conventional Dolbeault cohomologies on $X$ and $S^1/\bbZ$. In particular it means we can decompose de Rham cohomology groups on $M$ into $\delb'$ cohomology groups and since $\dd(\sigma\wedge\omega)$ is trivial in de Rham cohomology we have 
\begin{equation}
    [ \CF ] = [ F ] \in H^{[2,2]}_{\delb'}(M) ,
\end{equation}
so that the deformation problem is controlled by the physical flux. In the following calculations we will continue to use $\CF$ with the understanding that it represents the real four-form flux. The problem then reduces to calculating the $\delb'$ cohomologies in terms of the de Rham and Dolbeault cohomology groups on $X$ and $S^1/\bbZ_2$. Since $X$ is topologically a Calabi--Yau space we immediately have 
\begin{align}
\label{eq:CYconds}
H_{\delb}^{(0,1)}(X)\cong H_{\delb}^{(0,2)}(X) \cong H_{\delb}^{(1,0)}(X)\cong H_{\delb}^{(2,0)}(X)  =0
\end{align}
matching the assumption in the heterotic moduli papers~\cite{delaOssa:2015maa,delaOssa:2014cia,Anderson:2014xha}.

%

\subsubsection{The zeroth page}

Because the space $X\times S^1/\bbZ_2$ has boundaries, the first question in calculating the cohomology is to consider the boundary conditions on sections of the spaces that appear in the double complex in figure~\ref{fig:moduli_double_complex}.  We focus on the $\Lambda^{[p,q]}$ spaces, but the same discussion applies to $T^{[1,0]}\otimes\Lambda^{[p,q]}$ and $\wedge^3T^{[1,0]}\otimes\Lambda^{[p,q]}$. 

The simplest boundary conditions come from imposing that the section $\theta\in\Gamma(\Lambda^{[p,q]})$ is either odd or even under the $\bbZ_2$ action on $S^1$. Using the identification~\eqref{eq:Lambdapq-def}, we can write a given form as 
\begin{equation}
    \theta = \mu + \dd y \wedge \nu
\end{equation}
where $\mu$ and $\nu$ are sections of $\Lambda^{(p,q)}$ and $\Lambda^{(p.q-1)}$ respectively. Imposing that a section is odd/even then enforces the standard boundary conditions
\begin{equation}
\label{eq:odd/even}
\begin{aligned}
    \text{even $\theta$:} && && &&
    \left.\del_y\mu\right|_{X_{(i)}} &= 0 , & &&
    \left.\nu\right|_{X_{(i)}} = 0 , \\
    \text{odd $\theta$:} && && &&
    \left.\mu\right|_{X_{(i)}} &= 0 , & &&
    \left.\del_y\nu\right|_{X_{(i)}} = 0 , \\
\end{aligned}
\end{equation}
where $X_{(i)}$ with $i=1,2$ are the two boundaries of $M$ at either end of the $S^1/\bbZ_2$ interval.  The boundary sources in heterotic M-theory appear for odd fields $\theta_{\text{odd}}\in\Gamma(\Lambda^{[p,q]})$, and modify the boundary conditions on the $\mu_{\text{odd}}\in\Gamma(\Lambda^{(p,q)})$ component to
\begin{equation}
\label{eq:generic_boundary_conditions}
    \left.\mu_{\text{odd}}\right|_{X_{1}} = \gamma_{(1)}, \qquad
    \left.\mu_{\text{odd}}\right|_{X_{2}} = -\gamma_{(2)} , 
\end{equation}
for some fixed forms $\gamma_{(i)}$ on the boundaries (the minus sign for $\gamma_{(2)}$ is conventional). One can deal with these conditions by extracting a linear interpolating mode so that
\begin{align}\label{eq:generic_particular_and_free}
\theta & =\gamma_{(1)}-y\left(\gamma_{(1)}+\gamma_{(2)}\right)+\Delta_{\theta}
\end{align}
where $\Delta_{\theta}$ satisfies the standard odd boundary conditions given in~\eqref{eq:odd/even}. 

For the moduli in~\eqref{eq:type3-mod-fields}, $r$ and $\mu$ are even fields while $\kappa$ and $\theta$ are odd. Of the latter only the $\theta_{[1,2]}$ component is sourced, since, as we will see, the the deformations of the source $j_{(i)}$ that appear in~\eqref{eq:flux-sources} are of type $[1,2]$. These identifications mean that in the double complex we should identify 
\begin{equation}
    \begin{aligned}
    \ext^3T^{[1,0]}\otimes \Lambda^{[0,p]} & \quad \text{odd}, & && \qquad && 
    \Lambda^{[1,p]} & \quad \text{sourced}, \\
    T^{[1,0]}\otimes \Lambda^{[0,p]} & \quad \text{even}, & && &&
    \Lambda^{[3,p]} & \quad \text{even},
    \end{aligned}
\end{equation}
and also $\Lambda^{[0,3]}$ as odd.

\subsubsection{The First Page}

For the fields without boundary sources we can use the Kunneth theorem to calculate the $\delb'$ cohomologies. For example for $H_{\delb'}^{[p,q]}$, on $X\times S^1$ we have 
\begin{equation}
    \begin{aligned}
    H_{\delb'}^{[p,q]}(X\times S^1) & =\big(H_{\delb}^{(p,q)}(X)\otimes H^{0}_{\text{dR}}(S^{1},\bbC)\big)\\&
    \qquad\oplus\big(H_{\delb}^{(p,q-1)}(X)\otimes H^{1}_{\text{dR}}(S^{1},\bbC)\big)
    \end{aligned}
\end{equation}
If we restrict to the $\bbZ_{2}$-even part on $S^{1}$ only the $H^{0}_{\text{dR}}(S^{1},\bbC)\simeq\bbC$ component survives
\begin{equation}
\begin{aligned}
    H_{\delb'}^{[p,q]}(X\times S^1)^{\text{even}} 
    & \simeq H_{\delb}^{(p,q)}(X)\otimes H^{0}_{\text{dR}}(S^{1},\bbC) 
   \simeq H_{\delb}^{(p,q)}(X) .
\end{aligned}
\end{equation}
Conversely if we impose odd boundary conditions only the $H^{1}_{\text{dR}}(S^{1},\bbC)\simeq\bbC$ component survives 
\begin{equation}
\begin{aligned}
    H_{\delb'}^{[p,q]}(X\times S^1)^{\text{odd}} 
    & \simeq H_{\delb}^{(p,q-1)}(X)\otimes H^{1}_{\text{dR}}(S^{1},\bbC) 
   \simeq H_{\delb}^{(p,q-1)}(X) .
\end{aligned}
\end{equation}

We can define cohomology groups on $M=X\times S^1/\bbZ_2$ by taking only the classes which are even under the $\bbZ_2$ action, as these are single valued on the quotient $M=X\times S^1/\bbZ_2$.  This captures all the bulk moduli information, but not the localised boundary moduli.  
For a sourced $\bbZ_2$-odd field $\theta\in\Gamma(\Lambda^{[p,q]})$ with boundary sources as in~\eqref{eq:generic_boundary_conditions}, given the form~\eqref{eq:generic_particular_and_free}, $\delb'$-closure gives 
\begin{align}
\delb'\theta & =\delb\gamma_{(1)}-y(\delb\gamma_{(1)}+\delb\gamma_{(2)})-\dd y\wedge\left(\gamma_{(1)}+\gamma_{(2)}\right)+\delb'\Delta_{\theta}=0.
\end{align}
Since $\Delta_\theta$ is odd it needs to vanish on the boundaries, and so the general solution has $\Delta_\theta=-\dd y\wedge\eta+\tilde\Delta_\theta$ and 
\begin{equation}
\begin{aligned}
\delb\gamma_{(1)} & =0 ,\\
\delb\gamma_{(2)} & =0 ,\\
\delb\eta-\gamma_{(1)}-\gamma_{(2)} & =0 ,
\end{aligned}
\end{equation}
and $\delb'\tilde{\Delta}_\theta=0$. In particular we get the constraint that $[\gamma_{(1)}]=-[\gamma_{(2)}]$. One the other hand, modding out by $\delb'$-exact forms gives
\begin{equation}
    \theta \sim \theta + \delb'\nu = 
    \theta + \delb\epsilon_{(1)}-y(\delb\epsilon_{(1)}+\delb\epsilon_{(2)})-\dd y\wedge\left(\epsilon_{(1)}+\epsilon_{(2)}\right)+\delb'\Delta_{\nu} . 
\end{equation}
Correspond to $\gamma_{(i)}\sim\gamma_{(i)}+\delb\epsilon_{(i)}$ and $\eta\sim\eta+\epsilon_{(1)}+\epsilon_{(2)}$ together with $\tilde{\Delta}_\theta\sim\tilde{\Delta}_\theta+\delb'\Delta_\nu$. Since $\tilde{\Delta}_\theta$ is odd it is a representative of a class in $H^{[p,q]}(X\times S^1)^{\text{odd}}\cong H^{(p,q-1)}(X)$, which can equivalently be thought of as the ambiguity in determining $\eta$.  Combining this with the boundary terms, where $[\gamma_{(1)}]=-[\gamma_{(2)}]$ gives a single class in the group $H_{\delb}^{(p,q)}(X)$ we see that $\theta$ represents an element in
\begin{align}
H_{\delb'}^{[p,q]}(M)^{\text{sourced}} & \simeq H_{\delb}^{(p,q)}(X)\oplus H_{\delb}^{(p,q-1)}(X).
\end{align}
We can now identify the groups that appear in the moduli space calculation. First we recall that the flux is an element of $H_{\delb'}^{[2,2]}$ and is sourced by $j_{(i)}$. We have 
\begin{equation}
    [\CF] = [F] = [j_{(1)}] = - [j_{(2)}]
        \in H_{\delb}^{(2,2)}(X) 
        \subset H_{\delb'}^{[2,2]}(M)^{\text{source}} ,
\end{equation}
where we can write a real representative as
\begin{equation}
\label{eq:Flinear}
    \CF = F = j_{(1)} - y (j_{(1)}+j_{(2)}) - \dd y \wedge h ,
\end{equation}
where $\delb h=j_{(1)}+j_{(2)}$. The condition $[j_{(1)}] = - [j_{(2)}]$ reflects the fact that the total source $\rho_M$ in modified Bianchi identity $\dd F=\rho_M$ must, by definition, be trivial in cohomology. There is one moduli cohomology group absent from the spectral sequence as it doesn't interact with the other moduli,
\begin{align}
    H^{[0,3]}_{\delb'}(M) &\simeq H_{\delb}^{(0,2)}(X) = 0, && \text{odd} \\
\intertext{where we have used the Calabi--Yau result $H_{\delb}^{(0,2)}(X)=0$. The groups appearing on the first page are}
    H^p_{\delb'}(M,T^{[1,0]}) &\simeq H^p_{\delb'}(X,T^{(1,0)}) , && \text{even} \label{eq:T1X}\\
    H^{[1,p]}_{\delb'}(M) &\simeq H_{\delb}^{(1,p)}(X) 
       \oplus H_{\delb}^{(1,p-1)}(X) , && \text{sourced} \label{eq:oddX}\\
    H^{[3,p]}_{\delb'}(M) &\simeq H_{\delb}^{(3,p)}(X) , && \text{even} . \label{eq:evenX}
\end{align}


\subsubsection{The Second Page}

Moving to the second page, where we have to compute the cohomologies
of the $[\CF]$ map. Consider first 
\begin{equation}
    H^1_{\delb',\CF}(M,T^{[1,0]})  = 
        \ker \Bigl( H^{1}_{\delb'}(M,T^{[1,0]})\xrightarrow{\cdot[\CF]} H^{[1,3]}_{\delb'}(M) \Bigr) . 
\end{equation}
Let $r$ be a representative of $H_{\delb'}^{1}(M,T^{[1,0]})$ and $\CF$ be the  representative of the flux class given in~\eqref{eq:Flinear}. To lie in the kernel of the $[\CF]$ map we require $r\cdot\CF=\delb'\alpha_{[1,2]}$ for some $\alpha_{[1,2]}\in\Gamma\left(\Lambda^{[1,2]}\right)$ satisfying
$\alpha|_{X_{(1)}}=\tilde{\gamma}_{(1)}$ and $\alpha|_{X_{(2)}}=\tilde{\gamma}_{(2)}$ defined up to the addition of arbitrary $\delb'$-closed forms.  The kernel condition then reads 
\begin{equation}
\begin{aligned}
    r\cdot j_{(1)} &- y\left(r\cdot j_{(1)}+r\cdot j_{(2)}\right) - \dd y\wedge r\cdot h 
    \\ & \quad 
    =\delb\gamma_{(1)}-y\left(\delb\tilde{\gamma}_{(1)}+\delb\tilde{\gamma}_{(2)}\right)-\dd y\wedge\left(\tilde{\gamma}_{(1)}+\tilde{\gamma}_{(2)}\right) + \delb'\Delta_\alpha .
\end{aligned}
\end{equation}
Since $\delb'\CF=0$ implies $j_{(1)}+j_{(2)}=\delb h$ we immediately get the condition 
\begin{align}
\left[r\right]\cdot\left[j_{(1)}\right]\equiv\left[r\right]\cdot\left[j_{(2)}\right] & =0\in H_{\delb}^{(1,3)}\left(X\right)\label{eq:f_boundary_constraint}
\end{align}
It might appear that one needs to impose the further condition on $r$ that  $\left[\tilde{\gamma}_{(1)}+\tilde{\gamma}_{(2)}+r\cdot h\right]$ is trivial in $H_{\delb}^{(1,2)}(X)$. However $\tilde{\gamma}_{(i)}$ is defined up to the addition of an arbitrary $\delb$-closed form, meaning we may always shift the
$\tilde{\gamma}_{(i)}$ by a closed form such that the condition is satisfied. Thus we find that 
\begin{align}
    H^1_{\delb',\CF}(M,T^{[1,0]}) \simeq 
        \ker \Bigl( H^{1}_{\delb}(X,T^{(1,0)})\xrightarrow{\cdot[j_{(i)}]} H^{(1,3)}_{\delb}(X) \Bigr) \equiv H^1_{\delb,j_{(i)}}(X,T^{(1,0)}) . 
\end{align}
The next potentially non-trivial homology on the second page is
\begin{align}
    H^{[1,2]}_{\delb',\CF}(M)  = \frac{
        \ker \Bigl( H^{[1,2]}_{\delb'}(M)\xrightarrow{\wedge[\CF]} H^{[3,4]}_{\delb'}(M) \Bigr)}
        {\image\Bigl( H^{0}_{\delb'}(T^{[1,0]})\xrightarrow{\cdot[\CF]} H^{[1,2]}_{\delb'}(M) \Bigr)} .
\end{align}
However from~\eqref{eq:T1X} we see that $H_{\delb'}^{0}(T^{[1,0]})\simeq H_{\delb}^{0}(T^{(1,0)},X)\simeq H^{(2,0)}(X)=0$, where we use the Calabi--Yau conditions~\eqref{eq:CYconds}. Similarly from~\eqref{eq:evenX}, we have $H^{[3,4]}_{\delb'}(M)\simeq H^{(3,4)}_{\delb}(X)=0$. Hence we get 
\begin{align}
    H_{\delb',\CF}^{[1,2]}(M) \simeq H_{\delb'}^{[1,2]}(M) \simeq  H_{\delb'}^{(1,2)}(X) \oplus  H_{\delb'}^{(1,1)}(X) .  
\end{align}
We also immediately have 
\begin{equation}
\label{eq:Fcohomo-zeros}
    H_{\delb',\CF}^{0}(T^{[1,0]}) = 0 , \qquad
    H^{[3,4]}_{\delb',\CF}(M) = 0. 
\end{equation}
Finally we have
\begin{align}
H^{[3,3]}_{\delb',\CF}(M)  &= 
        \frac{H^{[3,3]}_{\delb'}(M)}{
        \image\Bigl( H^{[1,1]}_{\delb'}(M)\xrightarrow{\wedge[\CF]} H^{[3,3]}_{\delb'}(M) \Bigr).}
\end{align}
The quotient is trivial, since $H_{\delb}^{[3,3]}(M)$
is constructed from $\bbZ_{2}$-even forms on $S^1$, so is $H_{\delb'}^{[1,1]}(M)$
but $\CF$ is $\bbZ_{2}$-odd, so the $\left[\CF\right]$-map above must be trivial. We therefore simply have
\begin{align*}
H_{\delb',\CF}^{[3,3]}\left(M\right) & \simeq H^{[3,3]}_{\delb'}(M) \simeq H^{(3,3)}_{\delb}(X) . 
\end{align*}

\subsubsection{The Third Page}

Finally we have to consider the third page, where the non-trivial maps are
\begin{align}
\delta: H_{\delb',\CF}^{0}(M,T^{[1,0]}) \to H^{[3,3]}_{\delb',\CF}(M), \\
\delta': H_{\delb',\CF}^{1}(M,T^{[1,0]}) \to H^{[3,4]}_{\delb',\CF}(M) .
\end{align}
However from~\eqref{eq:Fcohomo-zeros} we see that actually both these maps are trivial and we get no further conditions. 


\subsubsection*{Heterotic Moduli}

In conclusion, we find that the heterotic M-theory moduli are  given by
\begin{equation}
\label{eq:final-modspace}
\begin{split}
\text{moduli space} & \simeq 
H_{\delb,j_{(i)}}^{1}(X,T^{(1,0)})
\oplus H_{\delb}^{(1,2)}(X) \oplus H_{\delb}^{(1,1)}(X)  \oplus H_{\delb}^{(3,3)}(X) .
\end{split}
\end{equation}
Group-by-group these cohomologies match those of the heterotic theory:
\begin{itemize}
    \item $H_{\delb,j_{(i)}}^{1}(X,T^{(1,0)})$ -- complex structure moduli, with some lifted by the flux sources $j_{(i)}$.  
    \item $H_{\delb}^{(1,2)}(X)$ -- moduli associated to deformations of the boundary sources. 
    \item $H_{\delb}^{(1,1)}(X)$ -- complex Kähler moduli.
    \item $H_{\delb}^{(3,3)}(X)$ -- axion-dilaton modulus.
\end{itemize}
These cohomology groups coincide with those found for the Hull--Strominger heterotic background in~\cite{Anderson:2014xha,delaOssa:2014cia,delaOssa:2015maa,hep-th/0509131,hep-th/0612290,Ashmore:2018ybe,Ashmore:2019rkx} under the assumption that $X$ is topologically a Calabi--Yau space (although the Hull--Strominger metric is not Ricci-flat).  To see the correspondence in detail, consider the form of the heterotic moduli problem given in~\cite{Ashmore:2019rkx}:
\begin{equation}
\label{eq:het-closed}
\begin{aligned}
\bar{\partial}\mu  =0, \\
\bar{\partial}b  =0, \\
\bar{\partial}x+2\mu\cdot h+2\tr(\alpha\wedge \mathcal{F})+\partial b  =0, \\
\bar{\partial}_{A}\alpha+\mu\cdot \mathcal{F}  =0 .
\end{aligned}
\end{equation}
where $\mu\in\Gamma(T^{(1,0)}\otimes\Lambda^{(0,1)})$, $b\in\Gamma(\Lambda^{(0,2)})$ and $x\in\Gamma(\Lambda^{(0,2)})$. The parameters  $\alpha\in\Gamma(\Lambda^{(0,1)}\otimes \ad P_G)$ encode deformations of the gauge (and gravity) fields $\mathcal{F}$ transforming in a total group $G$. Here it includes contributions from both boundaries. The object $h$ is the $(2,1)$-component of the heterotic three-form field strength $H$ and is related to the background $\omega$ by $h=\ii\del\omega$. The deformations are trivial if they have the form 
\begin{equation}
\label{eq:het-trivial}
\begin{aligned}
\mu&=-\bar{\partial}w, \\
x&=-\bar{\partial}\xi-\partial\bar{\xi}+2w\cdot h+2\tr(\theta\,\mathcal{F}), \\
b&=-\bar{\partial}\bar{\xi}, \\
\alpha&=-\bar{\partial}_{A}\theta+\imath_{w}\mathcal{F},  
\end{aligned}
\end{equation}
where $w\in\Gamma(T^{(1,0)})$, $\xi\in\Gamma(\Lambda^{(1,0)})$ and   $\theta\in\Gamma(\ad P_G)$. We can split the generic gauge fields $\mathcal{F}$ into parts that correspond to contributions from the two $E_8$ gauge fields on boundaries and the Riemann curvature, so that $\mathcal{F}=(\mathcal{F}_{(1)},\mathcal{F}_{(2)},\mathcal{R})$ and define 
\begin{align}
    j_{(i)} &= \tr\mathcal{F}^{(i)}\wedge\mathcal{F}^{(i)}-\tfrac{1}{2}\tr\mathcal{R}\wedge\mathcal{R}, \\
    \gamma_{(i)} &=\tr\big(\alpha^{(i)}\wedge\mathcal{F}^{(i)}\big), \\
    \nu_{(i)} &=\tr\big(\theta^{(i)}\wedge\mathcal{F}^{(i)}\big). 
\end{align} 
The heterotic Bianchi identity implies 
\begin{equation}
\label{eq:hetBI}
    \delb h = j_{(1)}+j_{(2)} , 
\end{equation}
so that $[j_{(1)}]=-[j_{(2)}]$ as above. (Note that the fact that $\gamma_{(i)}$ is of type $(1,2)$ is the reason we only included boundary sources for $\theta_{[1,2]}$ in the heterotic M-theory calculation.) If we assume $X$ is Calabi--Yau then we also have a $\del\delb$-lemma.  By taking traces of the last equations in~\eqref{eq:het-closed} and~\eqref{eq:het-trivial}, gauging away the terms with $\del$ by using the $\del\delb$-lemma, and noting $b$ is always trivial since $H^{(0,2)}(X)=0$ we get the closure conditions 
\begin{equation}
\label{eq:het-closed-simple}
\begin{aligned}
\bar{\partial}\mu & =0, \\
\bar{\partial}x+2\mu\cdot h+2(\gamma_{(1)} + \gamma_{(2)}) & =0, \\
\bar{\partial}\gamma_{(i)}+\mu\cdot j_{(i)} & =0 ,
\end{aligned}
\end{equation}
and the trivial variations 
\begin{equation}
\begin{aligned}
\mu&=-\bar{\partial}w, \\
x&=-\bar{\partial}\xi+2w\cdot h+2(\nu_{(1)}+\nu_{(2)}), \\
\gamma_{(i)} &=-\bar{\partial}\nu_{(i)} +\imath_{w}j_{(i)},  
\end{aligned}
\end{equation}
It is then easy to see that, given~\eqref{eq:hetBI}, that these equations have a moduli space that exactly matches~\eqref{eq:final-modspace}, with $\mu\cdot j_{(i)}$ being the only obstruction. 

In conclusion, we have shown that, under only the assumption that the supersymmetric involution condition holds, the heterotic M-theory and Hull-Strominger system moduli spaces agree.

\section{Conclusion and outlook}\label{sec:Conclusion and Outlook}
With this work we have completed the description of the torsion-free $\SU{7}$ generalised structures and their infinitesimal moduli for M-theory backgrounds.  Using type to partition our structures into those which admit non-trivial flux (type-3) and those which do not (type-0) we were then able understand how the constraints of involutivity translate to the underlying $\SU{3}$ or $\G_{2}$ structures.  We gave the general superpotentials for each type, showing that the vanishing of the superpotential and its first variations encompasses the involutivity conditions as expected.

The cohomology groups counting infinitesimal involutive moduli for a type-0 $\UniR{7}$ structure were detailed in \cite{Ashmore:2019rkx}. Under reasonable assumptions, we argued here that such backgrounds do not allow for sources and so the only compact examples are in fact $\G_2$ manifolds. We then also analysed the equations for the infinitesimal moduli fields for type-3 structures in the presence of nontrivial flux.  Unlike the type-0 case these cohomology groups are not de Rham or Dolbeault cohomologies in general, but instead are the cohomology groups of a generalised $\dd+\CF$ operator. We computed the $\dd+\CF$ cohomologies for the algebra of moduli fields using a spectral sequence which stabilised on the third page. (A similar analysis of the operators with non-trivial  NSNS three-form flux have appeared in~\cite{Rohm:1985jv,Cavalcanti:2005hq}.) The dimension of the cohomology groups weakly decreases as we ascend through the pages, implying that the number of moduli can only be reduced by the addition of flux. 

Computing the cohomology groups for a Hetertoric M-theory background $M=X\times (S^1/\bbZ_2)$ with $X$ a complex Calabi-Yau reproduced known results for dual Hull-Somtringer heterotic backgrounds, expanding on a linearised calculation given in~\cite{hep-th/9806051}. This calculation only relied on the complex involutive nature of the $\SU7$ structure, which appears as an F-term condition. Thus, if such terms are not corrected it has the potential to hold even when higher order corrections to the M-theory action are included. We also note that the corresponding superpotential for type-3 backgrounds, alongside its variations, could be used to compute 1-loop partition functions paralleling the recent work done in~\cite{Ashmore:2021pdm}.  This would allow further comparisons between M-theory and Heterotic backgrounds.  

The infinitesimal moduli computed here describe the tangent space to $\UniR{7}$-structure moduli space and we now have a clear picture of how this tangent space differs between type-0 and type-3 structures. It would of course be important to see if these are not obstructed once one goes away from the linear approximation. In addition, the moduli space of type-0 backgrounds contains within it the moduli space of torsion-free $\G_2$ structures, with the involutivity condition implying the $\Gx{2}$ structure is closed.  In particular, when there are no sources, these are the only solutions and so this formulation may have implications for understanding the existence of $\G_2$ geometries as well as the behaviour of a seven dimensional generalised Ricci flow. In upcoming work~\cite{ModuliStab}, we further extend this  analysis to type II backgrounds. In general one finds they match the na\"ive superpotential calculation, even when the background is far from a Calabi--Yau geometry. Again the calculation goes via a spectral sequence and the presence of flux can only `reduce the number of moduli as compared with the fluxless case. 

\acknowledgments

We thank Mariana Graña, Rahim Leung and especially David Tennyson for helpful discussions. GRS is supported by an EPSRC DTP studentship.   DW is supported in part by the STFC Consolidated Grant ST/T000791/1 and the EPSRC New Horizons Grant EP/V049089/1. We acknowledge the Mainz Institute for Theoretical Physics (MITP) of the Cluster of Excellence PRISMA+ (Project ID 39083149) and the University of California Irvine for hospitality and support during part of this work.

\appendix

\section{Type-3 as a limit of type-0}\label{app:type-3 as a limit of type-0}
In the standard killing spinor picture we use a function $\chi$ \eqref{def:L&S_chi_def} to quantify the difference in norms between the two killing spinors at each point.  The type-3 conditions is equivalent to the requirement that $\chi=0$ globally.  The condition that a $\UniR{7}$ structure is of type-3 is equivalent to the requirement that $\chi=0$ globally, so the killing spinors have the same norm globally.  In practice the easiest way to find $\psi$, which generates the $\SU{7}$ line bundle over the $\UniR{R}$ structure, for a type-3 background is to take the $\chi\to 0$ limit of the type-0 counterpart.  Including the $\chi$ dependence explicitly we have
\begin{align}
    \psi(\chi) &= \ee^{3\Delta}\ee^{A+\tilde{A}}\ee^{(\cos\chi\Omega_- + \ii\Omega_- -\sin\chi\ii\sigma\wedge\omega)/\sin\chi}\cdot \sin\chi \\
    &= \ee^{3\Delta}\ee^{A+\tilde{A}}(\sin\chi + \cos\chi\Omega_- + \ii\Omega+ - \ii\sin\chi\sigma\wedge\omega + \\&\qquad \frac{1}{2}\frac{1}{\sin\chi}j(\cos\chi\Omega_- + \ii\Omega+ - \ii\sin\chi\sigma\wedge\omega)\wedge (\cos\chi\Omega_- + \ii\Omega+ - \ii\sin\chi\sigma\wedge\omega) + \dots )\nonumber
\end{align}
For a type-3 background we simply want to evaluate $\psi(0)$, however in our analysis we are intersted in deformations of $\psi$ which may introduce a type-0 component.  We can kill two birds with one stone by taking $\chi<<1$ and expanding about $\chi=0$
\begin{align}
    \psi(\chi) &= \ee^{3\Delta}\ee^{A+\tilde{A}}\bigg(\chi + \Omega - \ii\chi\sigma\wedge\omega  \\&\quad + \frac{1}{2}\frac{1}{\chi} j\left(\left(1-\frac{1}{2}\chi^2\right)\Omega_+ + \ii\Omega_- - \ii\chi\sigma\wedge\omega\right)\wedge\left(\left(1-\frac{1}{2}\chi^2\right)\Omega_+ + \ii\Omega_- - \ii\chi\sigma\wedge\omega\right) + \dots \bigg)\nonumber\\
    &= \ee^{3\Delta}\ee^{A+\tilde{A}}(\Omega + \frac{1}{2}(\ii j(\sigma\wedge\omega)\wedge\Omega-\ii j \Omega\wedge\sigma\wedge\omega)  \\&\qquad +\chi(1 - \ii\sigma\wedge\omega - \frac{\ii}{4}j\Omega_+\wedge\Omega_- -\frac{\ii}{4}j\Omega_-\wedge\Omega_+ -\ii j(\sigma\wedge\omega)\wedge\sigma\wedge\omega ) +\dots  \nonumber \\
    &= \ee^{3\Delta}\ee^{A+\tilde{A}}\ee^{-\ii\sigma\wedge\omega}\left(\Omega + \chi\left(1 - \frac{1}{8}j\Omega\wedge\Bar{\Omega}-\frac{1}{8}j\bar{\Omega}\wedge\Omega \right) + \dots \right).
\end{align}
Finally arriving at the small $\chi$ expression
\begin{equation}
    \psi(\chi)= \ee^{3\Delta}\ee^{A+\tilde{A}}\ee^{-\ii\sigma\wedge\omega}\ee^{-\frac{1}{8}\Omega\wedge\bar{\Omega}}\cdot (\Omega + \chi)
\end{equation}
from which we can read off $\psi(0)$ for a type-3 structure and see how perturbations into type-0 structure will arise.  We could also have done this analysis $+3\ii$-eigenbundle which uniquely specifies the $\UniR{7}$ structure.  By using
\begin{equation}
    \cot\chi\Omega_+ + \csc\chi\Omega_- = \frac{1}{2\sin\chi} \left((\cos\chi+1)\Omega + (\cos\chi-1)\bar{\Omega}\right)
\end{equation}
we can write a generic section $V$ of a type-0 $L_3$ bundle as
\begin{equation}\label{type-0-to-type-3-sections}
    V \simeq \exp \left(\frac{1}{2\sin\chi}\big( (\cos\chi+1)\Omega + (\cos\chi-1)\bar{\Omega}\big)- i\sigma\wedge\omega\right) \cdot v
\end{equation}
for any $v\in T_\bbC M$. Since $[\sigma\wedge\omega, \Omega]=[\sigma\wedge\omega,\bar{\Omega}]=0$ we can split the exponent and expand only the $\Omega$ and $\bar{\Omega}$ part of the twisting factor
\begin{equation}
    V \simeq \exp(i\sigma\wedge\omega)\cdot  \bigg(v + \frac{\cos\chi+1}{2\sin\chi}v\lrcorner \Omega + \frac{\cos\chi-1}{2\sin\chi}v\lrcorner\overline{\Omega}+\frac{1}{8}\Big(\overline{\Omega}\wedge (v\lrcorner\Omega) + \Omega\wedge(v\lrcorner\overline{\Omega})\Big)\bigg)
\end{equation}
we again wish to take the $\chi=0$ value for this expression, and again it is illuminating to do a small $\chi$ expansion
\begin{equation}
    V \simeq \exp(i\sigma\wedge\omega)\cdot  \bigg(v + \left(\frac{1}{\chi}-\frac{\chi}{4}\right) v\lrcorner \Omega - \left(\frac{\chi}{4}\right)v\lrcorner\overline{\Omega}+\frac{1}{8}\Big(\overline{\Omega}\wedge (v\lrcorner\Omega) + \Omega\wedge(v\lrcorner\overline{\Omega})\Big)\bigg)
\end{equation}
We saw above that $\Omega$ can be used to define a rank $4$ distribution  
\begin{equation*}
    T^{[0,1]} = \ker \Omega = \{v\in T_\bbC: v\lrcorner \Omega = 0\}
\end{equation*} 
We can express any vector $v\in T_\bbC M$ as a sum of a part $x$ in $T^{[0,1]}$ and a part $y$ in the compliment $T_\bbC M/T^{[0,1]}$.  This decomposition is not unique, but the expression for $V$ is,
\begin{equation}
    V \simeq \exp(i\sigma\wedge\omega)\cdot  \bigg(x + y + \left(\frac{1}{\chi}-\frac{\chi}{4}\right) y\lrcorner \Omega - \left(\frac{\chi}{4}\right)x\lrcorner\overline{\Omega}+\frac{1}{8}\Big(\overline{\Omega}\wedge (y\lrcorner\Omega) + \Omega\wedge(x\lrcorner\overline{\Omega})\Big)\bigg).
\end{equation}
up to $\mathcal{O}(\chi^2)$.  We see that it is the $y$ component of $v$ which is responsible for the singular behaviour of $V$ in the $\chi\to 0$ limit.  This singular behaviour is not representative of some breakdown of our bundle, it simply represents the fact that as $\chi\to0$ our $L_3$ will no longer be the twist of $T_\bbC M$.  Any $V$ where $y=0$ will remain finite in the $\chi\to 0$ limit, and to see where these degrees of freedom reappear we choose a family of $v(\chi)= x+\chi y$ so that the $y$ component vanishes when $\chi=0$. Again up to $\mathcal{O}(\chi^2)$ the generalised vector $V(\chi)$ is then
\begin{equation}
    V(\chi)\simeq \exp(-i\sigma\wedge\omega)\cdot \bigg(x+\chi y + y\lrcorner \Omega- \left(\frac{\chi}{4}\right)x\lrcorner\overline{\Omega} +\frac{1}{8}\Big(\chi\overline{\Omega}\wedge (y\lrcorner\Omega) + \Omega\wedge(x\lrcorner\overline{\Omega})\Big)\bigg).
\end{equation}
In the $\chi\to 0$ limit the degrees of freedom previously encoded in $y$ are now encoded in a free 2-form $y\lrcorner \Omega$, giving
\begin{equation}\label{type-3-limit-of-type-0}
    L_{3} = \exp \bigg(-i\sigma\wedge\omega -\frac{1}{8}\Omega\wedge\overline{\Omega}\bigg)\Big(T_\bbC^{[0,1]}M + \Lambda^{(2,0)}T_\bbC^*M\Big).
\end{equation}



\bibliographystyle{JHEP}
\bibliography{References.bib}

\end{document}